\documentclass[amssymb,amsmath,12pt]{revtex4}
\pdfoutput=1
\usepackage{graphicx,epsfig}
\usepackage{color}
\def\beq{\begin{equation}}
\def\eeq{\end{equation}}
\def\be{\begin{equation}}
\def\ee{\end{equation}}
\def\bea{\begin{eqnarray}}
\def\eea{\end{eqnarray}}

\newcommand{\gsim}{\lower.7ex\hbox{$\;\stackrel{\textstyle>}{\sim}\;$}}
\newcommand{\lsim}{\lower.7ex\hbox{$\;\stackrel{\textstyle<}{\sim}\;$}}

\begin{document}


 \title{Detecting supernovae neutrino with Earth matter effect}
 \author{ Wei Liao}
 \affiliation{
  Institute of Modern Physics, School of Sciences, \\
 East China University of Science and Technology, \\
 130 Meilong Road, Shanghai 200237, P.R. China %
}


\begin{abstract}
 We study Earth matter effect in oscillation of supernovae neutrinos. We show that detecting 
 Earth matter effect gives an independent measurement of spectra of supernovae neutrinos, i.e. the 
 flavor difference of the spectra of supernovae neutrinos. We study the effect of energy resolution
 and angular resolution of final electron or positron on detecting the signal of Earth matter effect. 
 We show that varying the widths of energy bins in analysis can change the signal strength of Earth 
 matter effect and the statistical fluctuation. A reasonable choice of energy bins can both suppress 
 the statistical fluctuation and make out a good signal strength relative to the statistical fluctuation.  
 Neutrino detectors with good energy  resolution and good angular resolution are therefore preferred 
 so that there are more freedom to vary energy bins and to optimize the signal of Earth matter effect 
 in analyzing events of  supernovae neutrinos.

\end{abstract}
 \maketitle


 \section{ Introduction}
 As a direct consequence of supernovae(SNe) explosion,  SNe neutrinos carry not only a lot of energy of
 the explosion out of SNe but also a lot of information of the explosion, e.g. luminosity of
 the explosion, time scale of the explosion, phases of the explosion such as the accretion, the cooling,  etc.
 Moreover, different types of SNe or different explosion mechanisms of SNe can have different predictions on neutrino spectrum.
 So measuring neutrinos from SNe is important for understanding the SNe explosion.
 Not just about the explosion, SNe neutrinos, specifically the electron (anti-)neutrinos($\nu_e$ and ${\bar \nu}_e$) also 
 deeply involve into the rapid process of the nucleosynthesis of elements heavier than iron, and hence carry important information of 
 nucleosynthesis happened in SNe.
For these reasons,  measuring the properties of SNe neutrinos, e.g. the temperature, the time dependence, the spectrum and the flavor content etc.,
becomes one of the most important goals in the research of astro-particle physics after the discovery of neutrinos from SN 1987A ~\cite{1987A}.

There are several neutrino experiments under construction or under proposal which can possibly detect a burst of SNe neutrinos
from a source not far away from the Earth.  These experiments have different detection channels of neutrinos
which include inverse $\beta$ decay(IBD) with proton that detects  ${\bar \nu}_e$,  
IBD process with some heavy elements, e.g.  $^{12}$C, $^{40}$Ar etc,  that can detect $\nu_e$
or both of $\nu_e$ and ${\bar \nu}_e$,  
elastic scattering of neutrino with electron that detects a weighted sum of the fluxes of all flavors of neutrinos, 
elastic scattering of neutrino with proton or other heavy elements that detect a direct sum of the fluxes of
all flavors of neutrinos via the neutral current interaction.  If a SN burst is detected,  one may hope that 
all independent flavors of SN neutrinos can be measured to a good precision with all these possible detection channels.
Unfortunately, all these experiments are sensitive to some specific processes and hence sensitive to some 
specific flavors of SNe neutrinos.  They all have difficulties to measure all possible spectra of SNe neutrinos
by themselves.

In this article we study how the Earth matter effect in neutrino oscillation can help to make an independent measurement of 
the spectra of SNe neutrinos and to resolve the spectra of different flavors of neutrinos.
In the following we first make a quick review of SNe neutrinos and its flavor conversion inside
SNe for a typical model of core-collapse SNe, and discuss different approaches and experiments to measure them.
We will discuss detection schemes, in particular, in liquid scintillator(LS) detector used in JUNO experiment~\cite{JUNO} 
and in water-based liquid scintillator detector proposed in Jinping underground laboratory~\cite{JinpingNeu}.
Then we discuss the Earth matter effect in neutrino oscillation on the spectrum of SNe neutrinos. We discuss 
how to discriminate Earth matter effect through a typical IBD process.
Compared to previous works on the subject of Earth matter effect in SNe neutrinos, we 
analyze in particular the effect of energy resolution and angular resolution on measuring the Earth matter effect in
oscillation of SNe neutrinos, which has not been done in previous works.
In this article, we discuss the detection of SNe neutrinos using, as an example, the neutrino spectra from
the standard core-collapse SN~\cite{SNe}.  Detection of neutrinos in situations in other SNe or other models of SNe can be similarly discussed.

\section{Supernovae neutrino and its detection}
In standard scenario of core-collapse SNe~\cite{SNe},  a SN goes through three stages or phases, i.e. the infall, the accretion and the 
cooling phases. Neutrinos in these three phases have some different properties.  For example, in the infall phase($\lsim$ tens ms), 
$\nu_e$ is the dominant flavor of neutrino produced in this stage, and in the cooling phase($\gsim$ 0.5 s), all flavors of neutrinos
have similar luminosities and temperatures.  In the accretion phase,  all flavors of neutrinos can be significantly produced, and
significant differences in luminosity and temperature among different flavors of neutrinos can also exist in this phase.

A common feature in all these phases is that the temperatures are all limited to be less than tens MeV. Hence,
the energies of reactions inside the core-collapse SNe and energies of neutrinos produced inside SNe are limited 
to be less than several tens of MeV.
As a consequence, the muon (anti)neutrino($\nu_\mu$ and ${\bar \nu}_\mu$) and the tau (anti)neutrino
($\nu_\tau$ and ${\bar \nu}_\tau$)  can not be produced via charged current interaction in SNe.
These neutrinos can only be produced by the neutral current interaction, which is an interaction universal to all flavors of active neutrinos,
and they have the same initial temperatures and luminosities: 
$T_{\nu_\mu}=T_{{\bar \nu}_\mu}=T_{\nu_\tau}=T_{{\bar \nu}_\tau}=T_{\nu_X}=T_{{\bar \nu}_X}$,
$L_{\nu_\mu}=L_{{\bar \nu}_\mu}=L_{\nu_\tau}=L_{{\bar \nu}_\tau}=L_{\nu_X}=L_{{\bar \nu}_X}$.
In short, there are three independent initial spectra, luminosities and temperatures  of SNe neutrinos 
for $\nu_e$, ${\bar \nu}_e$ and $\nu_X$(${\bar \nu}_X$) separately,  i.e. $L_{\nu_e}$, $L_{{\bar \nu}_e}$, $L_{\nu_X}$ ($L_{{\bar \nu}_X}$)
and $T_{\nu_e}$, $T_{{\bar \nu}_e}$, $T_{\nu_X}$ ($T_{{\bar \nu}_X}$).

Neutrinos coming out of SNe undergo flavor transformation. If not considering the
collective effects of oscillation~\cite{collectiveEffect}, 
the oscillation of neutrinos is understood by the matter effect in Mikheyev-Smirnov-Wolfenstein(MSW) mechanism~\cite{w,ms}.
If not considering the possible effects of shock wave inside SNe,  the matter effect in neutrino oscillation is well understood.
 The very high density of matter inside 
SNe makes $\nu_e$, when produced, effectively the heaviest neutrino of active neutrinos 
and  ${\bar \nu}_e$ effectively lightest anti-neutrino of active anti-neutrinos.
So, for normal hierarchy(NH) of neutrino mass,  $\nu_e$ coincides with the heaviest neutrino $\nu^m_3$ in matter and 
${\bar \nu}_e$ coincides with the lightest anti-neutrino ${\bar \nu}^m_1$ in matter.
For inverted hierarchy(IH) of neutrino mass, ${\nu_e}$  coincides with the heaviest neutrino $\nu^m_2$ in matter and 
${\bar \nu}_e$ coincides with the lightest anti-neutrino ${\bar \nu}^m_3$ in matter.
For the neutrino mixing angles so far measured~\cite{RPP}, in particular for the 1-3 mixing angle $\theta_{13}$ measured in recent years
~\cite{theta13-1, theta13-2,theta13-3},  non-adiabatic effect in matter effect is negligible, and neutrinos or anti-neutrinos in their mass eigenstates, 
when propagating out of SN, smoothly become their corresponding mass eigenstates in vacuum.
The final fluxes of neutrinos outside SN should be as follows~\cite{DS}:
\bea
F_{\nu_e}\approx F^0_{\nu_X},~~F_{{\bar \nu}_e}\approx\cos^2\theta_{12}F^0_{{\bar \nu}_e}+\sin^2\theta_{12} F^0_{{\bar \nu}_X}, \label{flux1}
\eea
for NH, and
\bea
F_{\nu_e}\approx \cos^2\theta_{12}F^0_{\nu_X}+\sin^2\theta_{12} F^0_{\nu_e},  ~~F_{{\bar \nu}_e}\approx F^0_{{\bar \nu}_X}, \label{flux2}
\eea
for IH.  In (\ref{flux1}) and (\ref{flux2}) ,  we have neglected the small correction at the order of $\sin^2\theta_{13}$ 
which is about $2\%$  according to the recent precise measurement of $\theta_{13}$~\cite{theta13-3,RPP}.
$F^0$ represents the initial fluxes of the corresponding species of neutrinos.
$F^0_{\nu_X}=F^0_{{\bar \nu}_X}$.
$\theta_{12}$ is the 1-2 mixing angle measured in the oscillation of solar neutrinos~\cite{RPP}.

In Water-Cherenkov(WC) detector or LS detector, major detection channel of SNe neutrinos is the IBD process with  proton
\bea
{\bar \nu}_e +p \to n + e^+ , \label{IBD1}
\eea
which probes the flux of ${\bar \nu}_e$ arriving at detector.
This process has a threshold energy of $1.8$ MeV.  For relativistic final electron, the cross section of this process is
approximately $ 0.67\times 10^{-43}$ (E$_\nu$/MeV)$^2$ cm$^2$ where $E_\nu$ is the energy of neutrino.  
Events of this process can be re-constructed event-by-event in WC detector or in LS detector.  
One of the problem for WC detector is that it does not have a very good energy resolution. 
The energy resolution in Super-Kamiokande(Super-K) detector is estimated as 
$\Delta_e/E_e =  (0.5 \sim 0.6)/\sqrt{\textrm{E/MeV}}$~\cite{SK-energyresolution} where $E_e$ is the energy of final $e^+$ in IBD process
and $\Delta_e$ is the energy uncertainty of $e^+$.  As will be discussed later, this energy resolution is not good
for detecting the Earth matter effect in oscillation of SNe neutrinos.

A great virtue of LS detector is that it has a very good
energy resolution. For the LS detector under construction in JUNO experiment, the energy resolution of electron or positron
is expected to reach $\Delta_e/E_e =  0.03/\sqrt{\textrm{E/MeV}}$~\cite{JUNO}.
For neutrinos with energy $\lsim 10$ MeV, the energy of initial neutrino in reaction
can also be re-constructed to such a high precision using the measured value of
$E_e$. For neutrinos with higher energy, the re-construction of neutrino energy to such a high precision requires a good angular
resolution of the final positron because $E_e$ and $E_\nu$ has a relation as follows
\bea
E_\nu=\frac{E_e+m_n-m_p}{1-\frac{E_e}{m_p}(1-\frac{|{\vec p}_e|}{E_e}\cos\theta_e )}, \label{relation}
\eea
where $m_p$ is the proton mass,  ${\vec p}_e$ the 3-momentum of $e^+$ and $\theta_e$  the angle between ${\vec p}_e$ and the momentum of neutrino.
One can read from (\ref{relation}) that without knowing the direction of $e^+$,
there is an extra uncertainty, of order $E_e/m_p$($\approx E_\nu/m_p$),  contributing 
to the re-construction of neutrino energy.  A nice advantage of water-based LS detector
is that it can achieve a high energy resolution as well as a good angular resolution~\cite{JinpingNeu}, 
so that a good resolution of neutrino energy after reconstruction can be achieved. 

In the following analysis,
we partially study the effect of a good  energy resolution of the reconstructed neutrino on analysing the Earth matter effect in oscillation
of SNe neutrinos. We are not going to concentrate on a specific assumption of energy resolution and angular resolution 
of a specific experiment. For simplicity, we make a linear sum of two uncertainties contributing to the re-constructed energy
of neutrinos. For example,  we can take the resolution of the re-constructed energy of neutrino as
\bea
\Delta_\nu/E_\nu= r_e/\sqrt{\textrm{E$_\nu/$MeV}}+r_a ~E_\nu/m_p, \label{ERES1}
\eea
where $r_e$ and $r_a$ are two numbers.  $r_a$ depends on the angular resolution of a specific experiment and can  vary from around 0.1 to around 1.
$r_e$ depends on the resolution of  energy of electron or positron.  
For practical analysis of neutrino events,  we also use $\Delta_\nu$ and (\ref{ERES1}) 
as the assumption of the width of energy bins of neutrinos in data analysis. That is, we can take
value of $r_e$ larger than the value of the designed energy resolution and hence take larger width of energy bins.
For example, for LS detector in JUNO experiment~\cite{JUNO}, we can take
$r_e > 0.03$ in (\ref{ERES1}) for a data analysis.  In later analysis in this article, we will use (\ref{ERES1}) 
as the assumption of the width of energy bins of neutrinos which is limited by the energy resolution and
angular resolution of electron or positron, but not the same.

The elastic scattering(ES) processes
\bea
\nu_e({\bar \nu}_e) +e \to \nu_e({\bar \nu}_e) +e, ~~\nu_x({\bar \nu}_x)+e\to \nu_x({\bar \nu}_x)+e,
\label{ES}
\eea
are also important detection channels of SNe neutrinos in WC and LS detectors. 
Since all flavors of neutrinos contribute to ES,
the ES process can not distinguish contributions of different flavors of neutrinos. Rather, it measures a
weighted sum of fluxes of $\nu_e$, ${\bar \nu}_e$ and $\nu_x$(${\bar \nu}_x$) because the
cross sections of these ES are not the same for $\nu_e$, ${\bar \nu}_e$ or $\nu_x$(${\bar \nu}_x$) and 
differ by a factor around $2\sim 7$. Comparing with the IBD process in (\ref{IBD1}), there are two major problems associated with this process.
First, the energy of neutrino can not be reconstructed 
from the energy of electron at a precision as good as in the IBD process because of the missing energy of final neutrino.
This feature makes this ES difficult to study carefully the spectrum of SNe neutrinos and the Earth matter effect in oscillation of SNe neutrinos. 
Second, the cross section of this process scales as $E_\nu$, not as fast as that of the IBD process,  and is smaller than that of the IBD process by a factor of
several tens to around a hundred for neutrinos with energy around tens of MeV.  So the event rate of ES is generally much smaller than
that of the IBD process in (\ref{IBD1}) even though all flavors of SNe neutrinos contribute to the scattering while only ${\bar \nu}_e$
contribute to this IBD process.  Another ES process $\nu +p \to \nu +p$ has a very small recoil energy of proton. So using this process is hard
 to detect SNe neutrinos.

For WC or LS detectors, there are also other reactions which can detect SNe neutrinos. For example, reactions with $^{12}$C
\bea
&& \nu + ~^{12}\textrm{C}\to \nu +~ ^{12}\textrm{C}^*,~~\label{reaction2-1}\\
&& \nu_e +~^{12}\textrm{C} \to e^- + ~ ^{12}\textrm{N},~~\label{reaction2-2}\\
&& {\bar \nu}_e +~^{12}\textrm{C} \to e^+ + ~ ^{12}\textrm{B} , \label{reaction2-3}
\eea
have three independent channels and can detect all flavors of neutrinos in JUNO~\cite{JUNO,JUNO2}. In particular, (\ref{reaction2-1})  and (\ref{reaction2-2})
can all offer measurements of neutrinos arriving at detector independent of those measurements by (\ref{IBD1}) and
(\ref{ES}). However, the event rates of these processes are
much smaller than that of the IBD process in (\ref{IBD1}).  In particular, the event rate of the reaction (\ref{reaction2-2})  which can detect
$\nu_e$ flavor, is around 50 to 100 times smaller than that of the IBD process in (\ref{IBD1})~\cite{JUNO}. Moreover, the energy of neutrino
of reaction (\ref{reaction2-1}) is hard to reconstruct so that this process is not really useful to reconstruct the spectrum of SNe neutrinos.
Similar problems exists in reactions with $^{16}$O, e.g. for WC detector.

Other neutrino detectors under proposal~\cite{EUproposal, LENA,DUNE} uses similar detection schemes discussed above or
Liquid Argon TPC(LaTPC) detector. A major advantage of LaTPC detector is that it is most sensitive to the $\nu_e$ flavor of neutrino.
Some analysis of the detection of SNe neutrinos using LaTPC has been done~\cite{LaTPC}.  The analysis performed in this article
concerning the Earth matter effect can be similarly applied to LaTPC detector or other detectors of low energy neutrinos
if they have a good resolution of the energy of reconstructed neutrinos.

As a short summary of this section,  major detectors of low energy neutrinos are all sensitive to some specific flavor of neutrinos.
Unless a SN is very close to the Earth so that the flux of SNe neutrinos arriving at the Earth is extremely high, these experiments
all have problems to measure completely all independent spectra of SNe neutrinos by themselves using the detection schemes
described above. As will be discussed below,
a measurement of the Earth matter effect in oscillation of SNe neutrinos offers an independent measurement of the spectra of SNe neutrinos
and can help to discriminate the flavor difference of neutrino spectra.

\section{Earth matter effect}
For SNe neutrinos passing through the Earth, Earth matter effect in neutrino oscillation can slightly change the flavor content of neutrinos.  
Since a burst of SN and a burst of SN neutrino lasts for about 10 s, the baseline of SNe neutrinos crossing the Earth can be considered fixed.
For neutrinos with maximally tens of MeV,  matter effect in 1-3 oscillation
is negligible~\cite{Liao, Liao1} and the Earth matter effect mainly affects 1-2 oscillation. 
Moreover, neutrinos from SNe should 
arrive at the Earth in mass eigenstates $\nu_1, \nu_2, \nu_3$ in vacuum. The effects of the Earth matter in neutrino oscillation
are encoded in the 
probabilities of $\nu_2 \to \nu_e$  and ${\bar \nu}_2 \to {\bar \nu}_e$:
\bea
P(\nu_2 \to \nu_e) \approx \sin^2\theta_{12}+f_{reg},~~P({\bar \nu}_2 \to {\bar \nu}_e) \approx \sin^2\theta_{12}+{\bar f}_{reg},
\label{reg}
\eea
where $f_{reg}$ and ${\bar f}_{reg}$, the regeneration factors, represent the effect of Earth matter on oscillation of
neutrinos and anti-neutrinos respectively.  In (\ref{reg}) we have neglected the small correction at the order of $\sin^2\theta_{13}$  
which is about $2\%$, as in (\ref{flux1}) and (\ref{flux2}).
$f_{reg}$ and ${\bar f}_{reg}$ are non-zero for up-going neutrino events.
For down-going neutrino events in detectors,  they are both zero.
In accord with (\ref{reg}),  we also have
$P(\nu_1 \to \nu_e) \approx \cos^2\theta_{12}-f_{reg}$ and $P({\bar \nu}_1 \to {\bar \nu}_e) \approx \cos^2\theta_{12}-{\bar f}_{reg}$.
For NH, the fluxes of $\nu_1$ and $\nu_2$ arriving at the Earth are all given by $F^0_{\nu_X}$,  and
the fluxes of ${\bar \nu}_1$ and ${\bar \nu}_2$ are given by $F^0_{{\bar \nu}_e}$ and $F^0_{{\bar \nu}_X}$
respectively~\cite{DS}. Using above equations of probabilities of  $\nu_i \to \nu_e$ and
${\bar \nu}_i \to {\bar \nu}_e$,  the final fluxes of $\nu_e$ and ${\bar \nu}_e$ are obtained as follows
\bea
F_{\nu_e}\approx F^0_{\nu_X},
~~F_{{\bar \nu}_e}\approx \cos^2\theta_{12}F^0_{{\bar \nu}_e}+\sin^2\theta_{12} F^0_{{\bar \nu}_X}
+{\bar f}_{reg} (F^0_{{\bar \nu}_X}-F^0_{{\bar \nu}_e}). \label{reg0a}
\eea
For IH,  the fluxes of $\nu_1$ and $\nu_2$ arriving at the Earth are given by $F^0_{\nu_X}$ and $F^0_{\nu_e}$
respectively, and the fluxes of ${\bar \nu}_1$ and ${\bar \nu}_2$  are all given by $F^0_{\nu_X}$.
Hence, the final fluxes of $\nu_e$ and ${\bar \nu}_e$ are
\bea
F_{\nu_e}\approx \cos^2\theta_{12}F^0_{\nu_X}+\sin^2\theta_{12} F^0_{\nu_e}+f_{reg} (F^0_{\nu_e}-F^0_{\nu_X}), 
 ~~F_{{\bar \nu}_e}\approx F^0_{{\bar \nu}_X}, \label{reg0b}
\eea
Again, in (\ref{reg0a}) and (\ref{reg0b}),  we have neglected the small correction at the order of $\sin^2\theta_{13}$  which is about $2\%$,
as in (\ref{flux1}) and (\ref{flux2}).
Apparently, if a measurement of Earth matter effect in SNe neutrinos can be performed, it would be
a measurement of the flavor-difference of the fluxes of neutrinos. For NH this means
$F^0_{{\bar \nu}_X}-F^0_{{\bar \nu}_e}$ in (\ref{reg0a}) , and for IH this means $F^0_{\nu_e}-F^0_{\nu_X}$ in (\ref{reg0b}).
This measurement of the spectrum of SNe neutrinos can not be done by the detecting processes described in
the last section and is complementary to those measurements.

\begin{figure}[tb]
\begin{center}
\begin{tabular}{cc}
\includegraphics[scale=1,width=8cm]{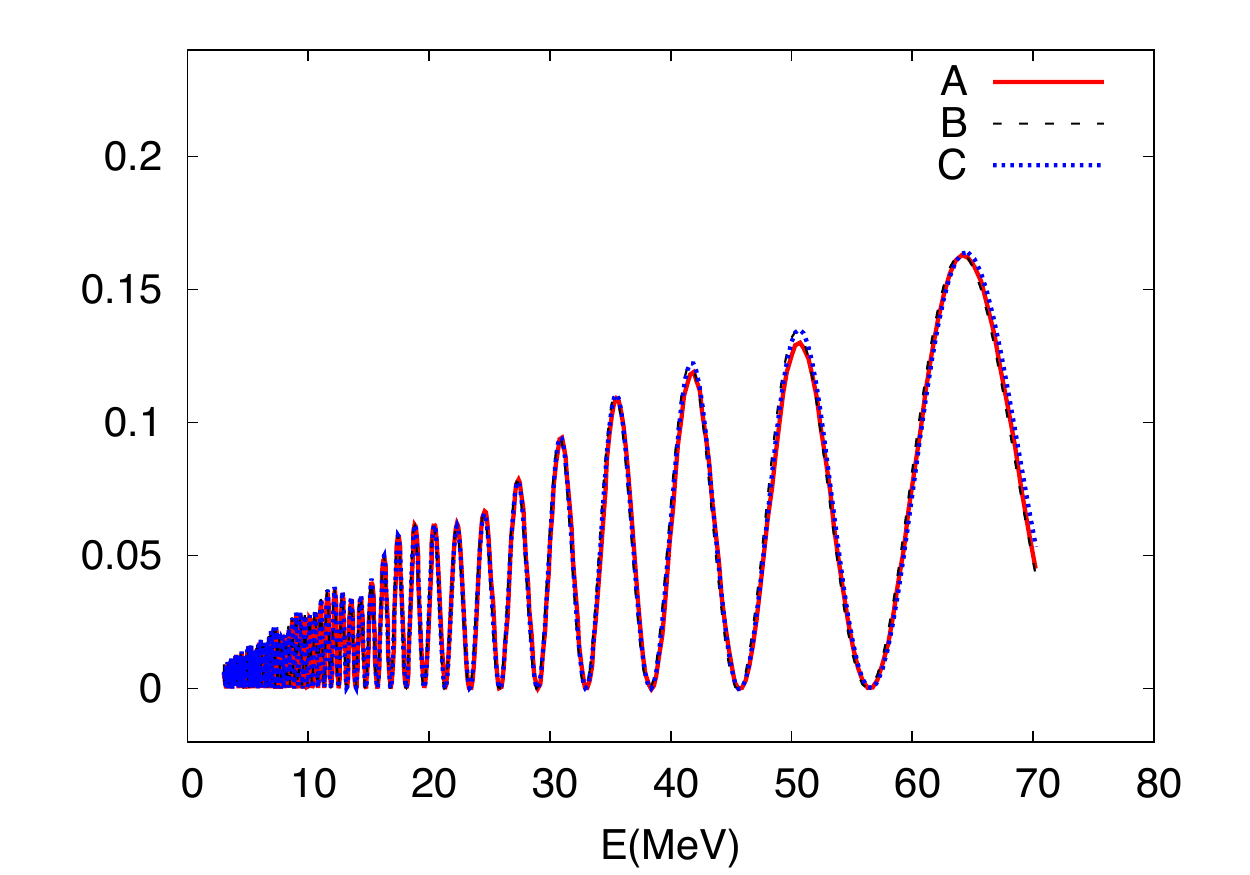}
\includegraphics[scale=1,width=8cm]{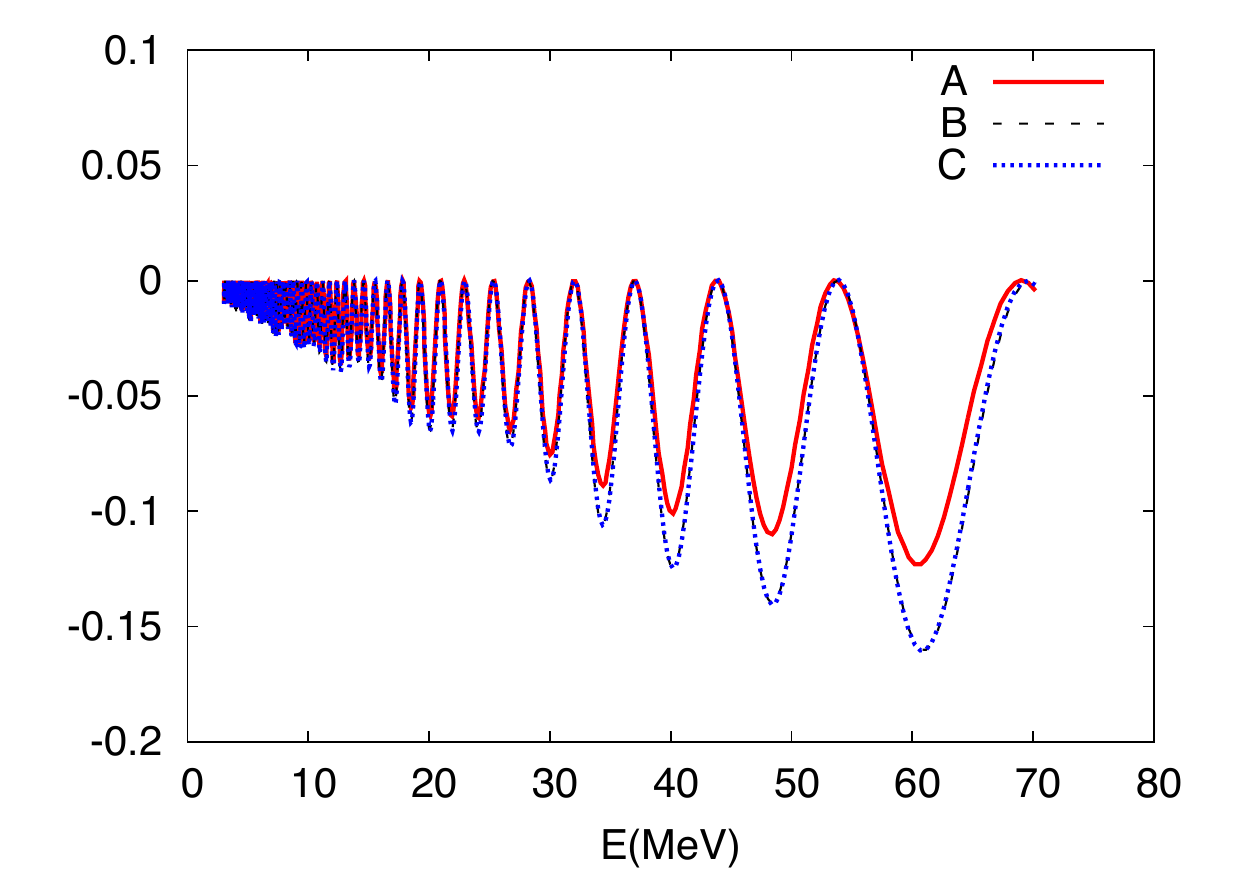}
\\
\includegraphics[scale=1,width=8cm]{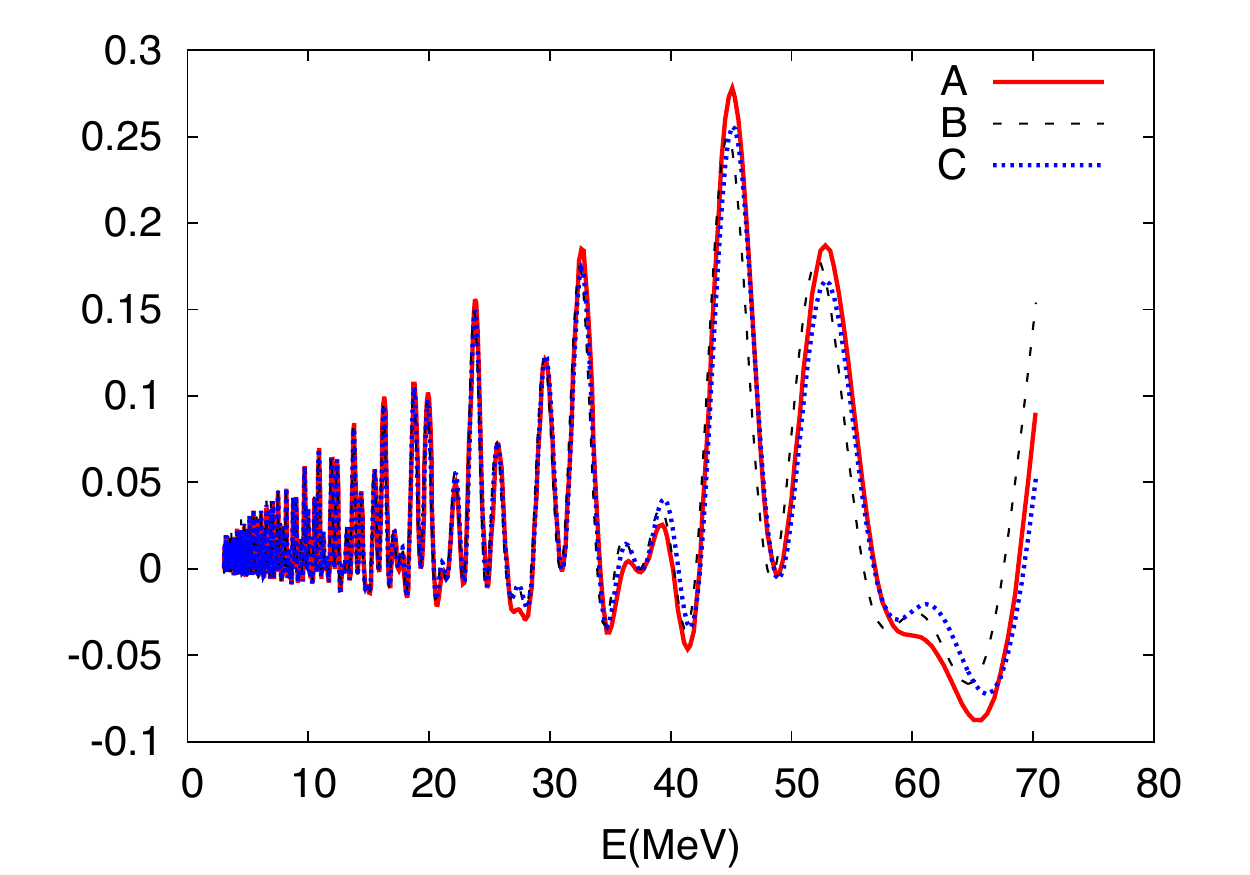}
\includegraphics[scale=1,width=8cm]{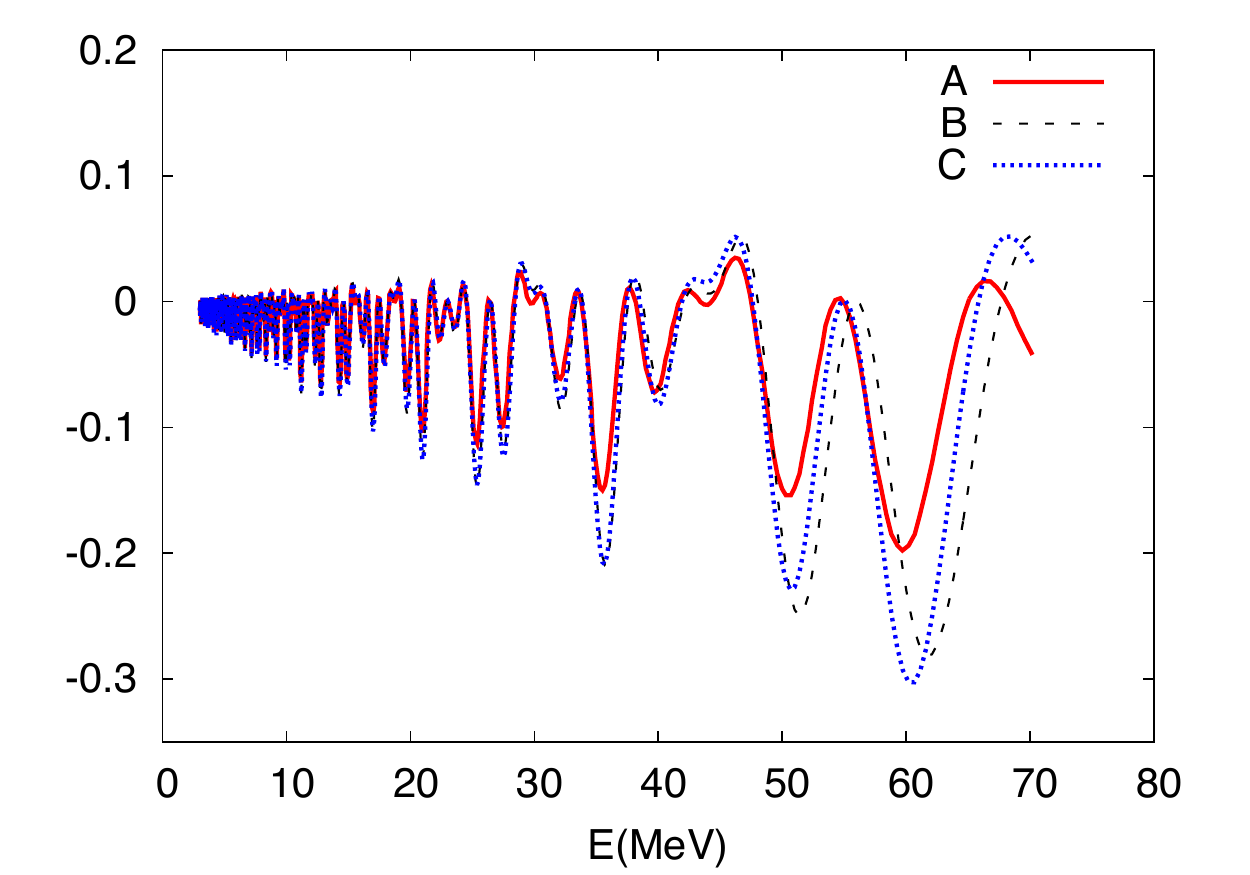}
\end{tabular}
\end{center}
\caption{ Regeneration factor versus energy for neutrino(Left) and anti-neutrino(Right) respectively.
$L=8000$ km (Upper) and $L=12000$ km (Lower) respectively. Line A(numerical): computed numerically
using PREM density profile with 5 layers; Line B(partial-analytical): computed using formula  (\ref{reg1a}) and (\ref{reg1b}) with
oscillation phase computed numerically; Line C(analytical): computed using formula  (\ref{reg1a}) and (\ref{reg1b}) 
completely analytically including the oscillation phase.}
\label{figure1}
\end{figure}

For neutrinos with energy around tens of MeV, the Earth matter effect in neutrino oscillation can be well described in an adiabatic perturbation theory
of oscillation~\cite{Liao0}.  In this theory, the regeneration factor due to Earth matter can be described by simple formulae as follows
\bea
&& f_{reg}=-\frac{E \sin^2\theta_{12}}{\Delta m^2_{21}} \sum_{i=0}^{k} \Delta V_i \cos2 \Phi_i,  \label{reg1a} \\
&& {\bar f}_{reg}=\frac{E \sin^2\theta_{12}}{\Delta m^2_{21}} \sum_{i=0}^{k} \Delta V_i \cos2 {\bar \Phi}_i,  \label{reg1b}
\eea
where $\Delta V_i$ is the potential jump at the i$^{th}$ density jump in-between layers of the Earth.  $\Phi_i$  and  ${\bar \Phi}_i$ are
the oscillation phases from the i$^{th}$ jump to detector for neutrino and anti-neutrino respectively.
For the symmetric PREM density profile of the Earth~\cite{PREM},  (\ref{reg1a}) and (\ref{reg1b}) can be re-written in a symmetric form~\cite{Liao0}.
It has been shown in \cite{Liao0} that (\ref{reg1a}) and (\ref{reg1b}) work very well for solar neutrinos. 
For application to SN neutrinos which have average energy slighter larger than solar neutrinos, detailed numerical analysis of
 the usefulness of (\ref{reg1a}) and (\ref{reg1b})  have not been given.  So we first analyze the usefulness of  (\ref{reg1a}) and (\ref{reg1b})
  for SN neutrinos. 
  
In Fig. \ref{figure1} we plot the regeneration factor versus energy for baselines of 8000 km and 12000 km respectively.
In this plot and thereafter in the present article we use~\cite{RPP}
\bea
\Delta m^2_{21}=7.5\times 10^{-5}~ {\textrm eV}^2, ~\sin^2 2\theta_{12}=0.857.
\eea
We compare the result calculated numerically using the PREM density profile, the result calculated analytically using the formulae
(\ref{reg1a}) and (\ref{reg1b}) , and the result calculated using  the formulae (\ref{reg1a}) and (\ref{reg1b}) but
with the oscillation phases calculated numerically(partial-analytical). For the analytical result, the oscillation phase can be calculated analytically using
an expansion in $E V_e/\Delta m^2_{21}$ of the oscillation phase~\cite{Liao0} and an approximate expression of the Earth density profile~\cite{PREM}.

\begin{figure}[tb]
\begin{center}
\begin{tabular}{cc}
\includegraphics[scale=1,width=8cm]{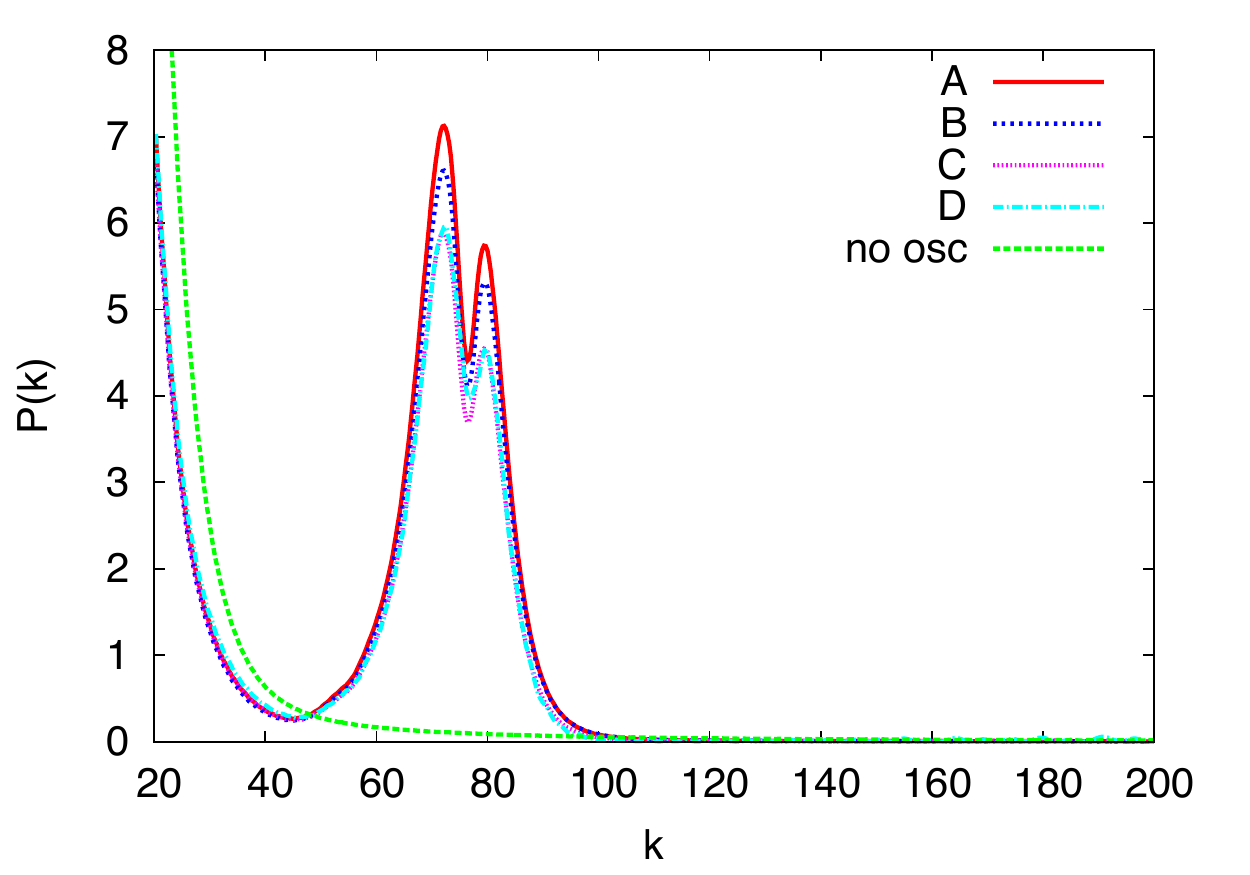}
\includegraphics[scale=1,width=8cm]{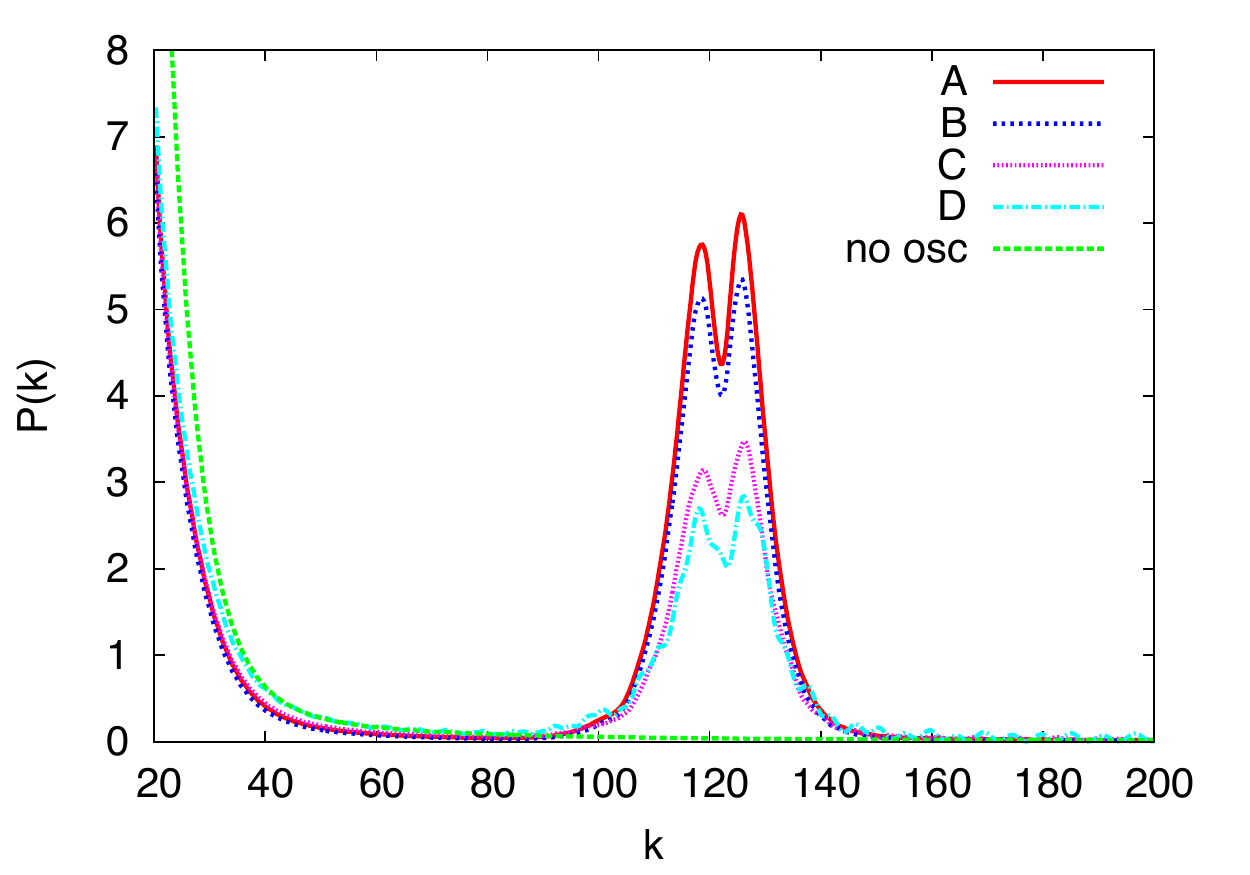}
\\
\includegraphics[scale=1,width=8cm]{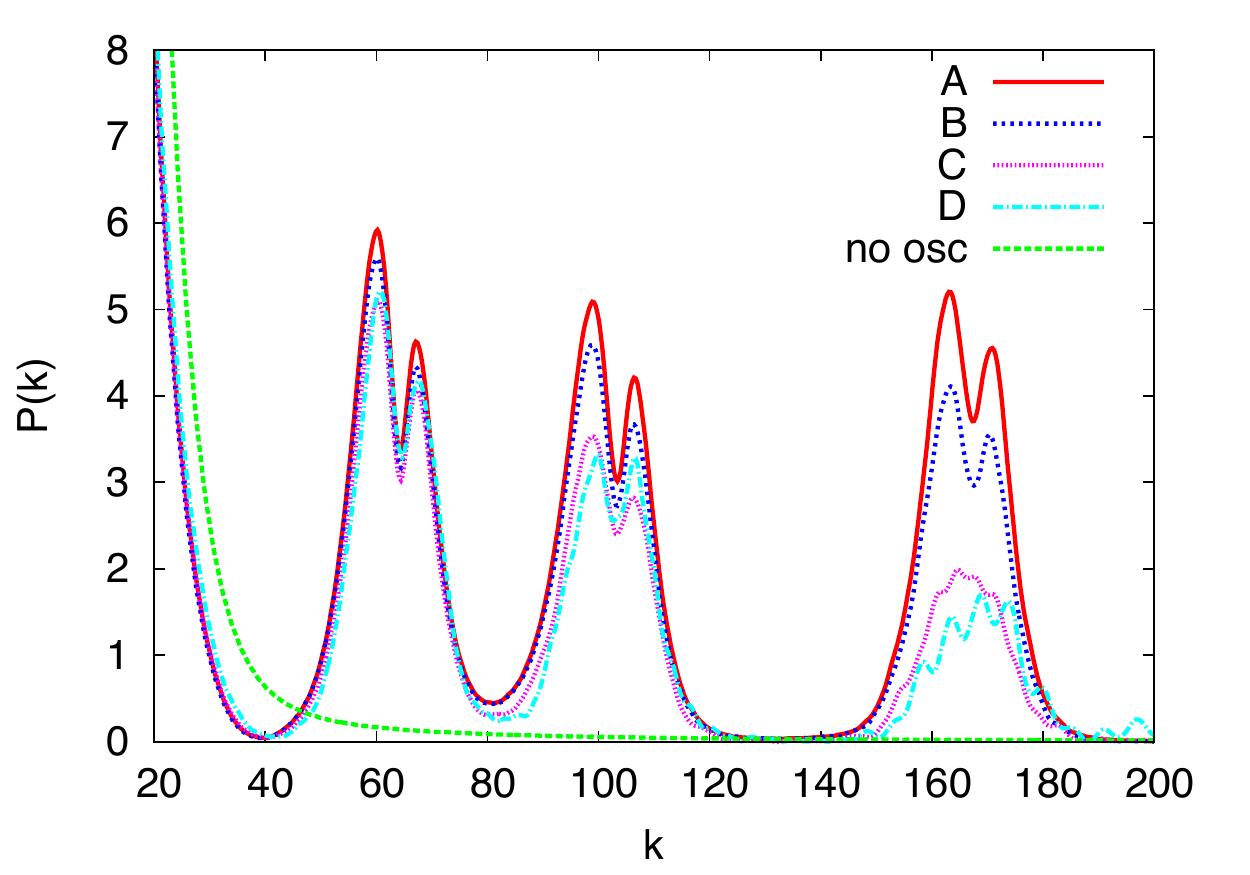}
\includegraphics[scale=1,width=8cm]{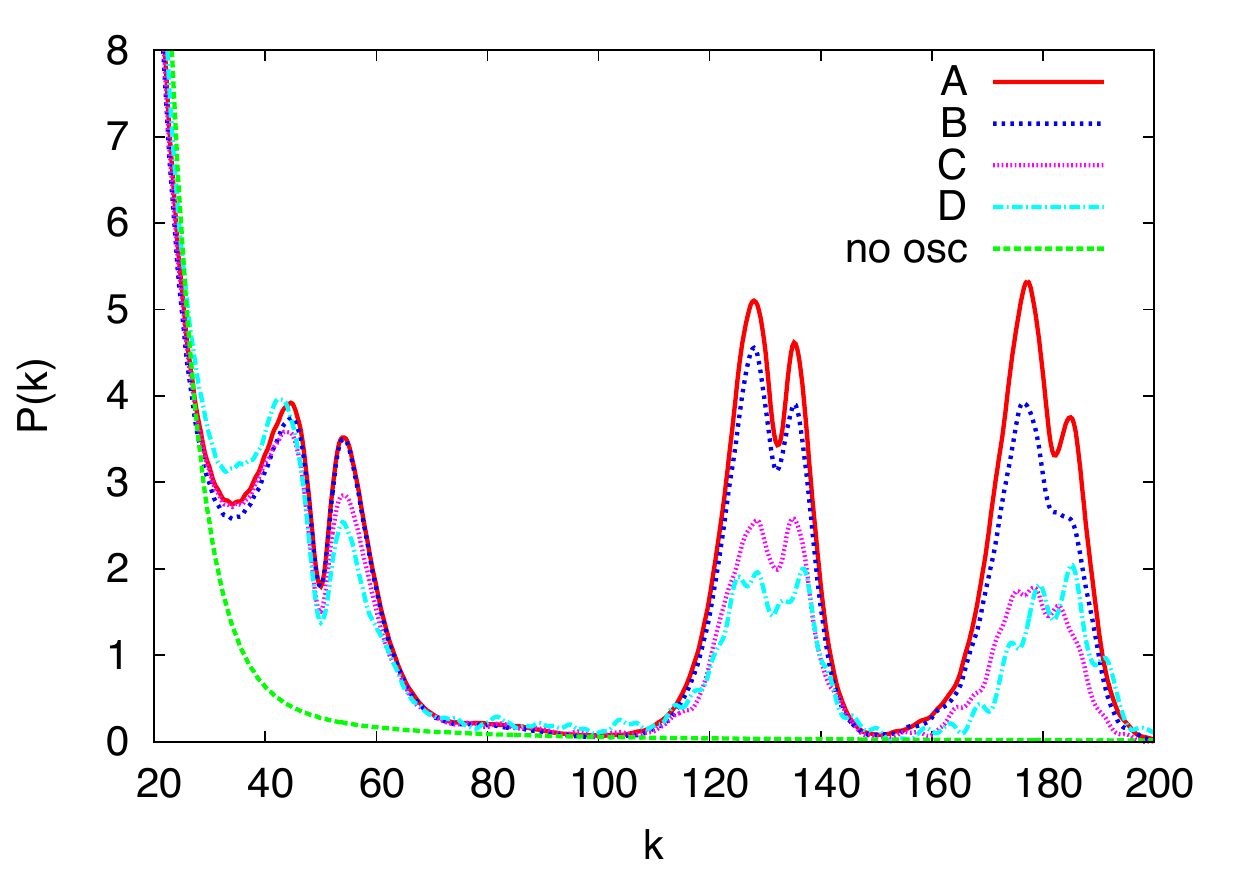}
\end{tabular}
\end{center}
\caption{ Power spectrum versus k for $L=5000$ km(Upper-Left), $L=8000$ km(Upper-Right),
$L=11000$ km(Lower-Left), $L=12000$ km(Lower-Right) with various assumptions of energy bin.  
Earth matter effect calculated numerically using PREM density profile with 5 layers.
Line A:  $\Delta_\nu/E_\nu=0.03/\sqrt{E_\nu/\textrm{MeV}}+0.2 \times E_\nu/m_p$;
Line B:  $\Delta_\nu/E_\nu=0.03/\sqrt{E_\nu/\textrm{MeV}}+0.5 \times E_\nu/m_p$;
Line C:  $\Delta_\nu/E_\nu=0.03/\sqrt{E_\nu/\textrm{MeV}}+1.0 \times E_\nu/m_p$;
Line D:  $\Delta_\nu/E_\nu=0.05/\sqrt{E_\nu/\textrm{MeV}}+1.0 \times E_\nu/m_p$.
No. of events: $2.\times 10^4$.}
\label{figure2}
\end{figure}

One can see in Fig. \ref{figure1} that the analytical and partial-analytical results both follow well the oscillation patterns.
For neutrino and baseline of 8000 km, there is no visible difference among numerical, analytical and partial-analytical results.
For neutrino and baseline of 12000 km, there are some visible differences among these three results when the energy is large(E $\gsim 60$ MeV).
But one can see in the plots that the differences are quite small.
For anti-neutrino and baseline of 8000 km, the difference between the analytical and partial-analytical results is not visible in the plot
but their difference to the numerical result becomes non-negligible for energy $\gsim 50$ MeV. 
For anti-neutrino and baseline of 12000 km, the difference between the analytical and partial-analytical results becomes visible in the plot
for energy $\gsim 50$ MeV and again the differences of these two results to the numerical result are non-negligible for energy
$\gsim 50$ MeV. As a short summary, one can see in these plots that the oscillation phase and oscillation pattern are well re-produced by 
the formulae (\ref{reg1a}) and (\ref{reg1b}) although the magnitude of oscillation is not re-produced very well for anti-neutrino
and for energy $\gsim 50$ MeV.

\begin{figure}[tb]
\begin{center}
\begin{tabular}{cc}
\includegraphics[scale=1,width=8cm]{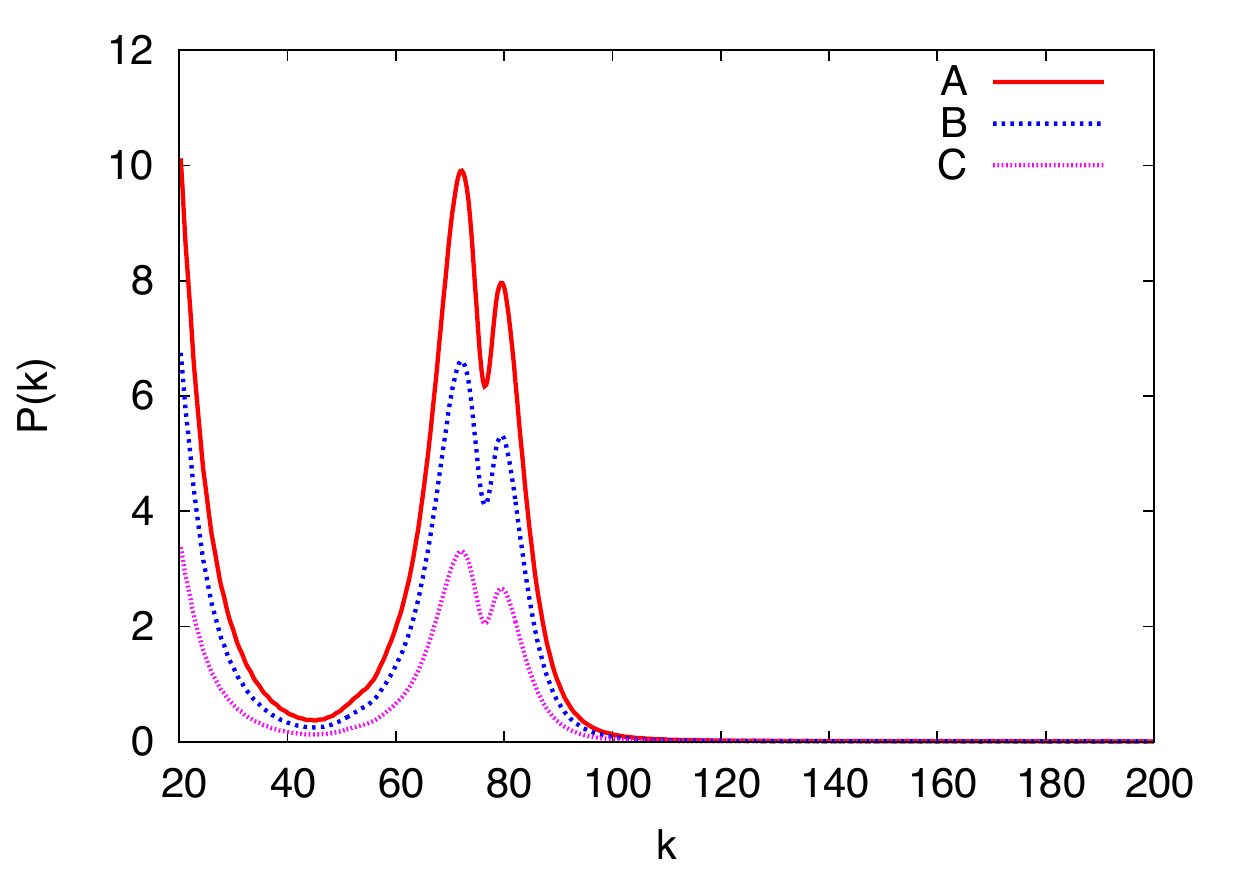}
\includegraphics[scale=1,width=8cm]{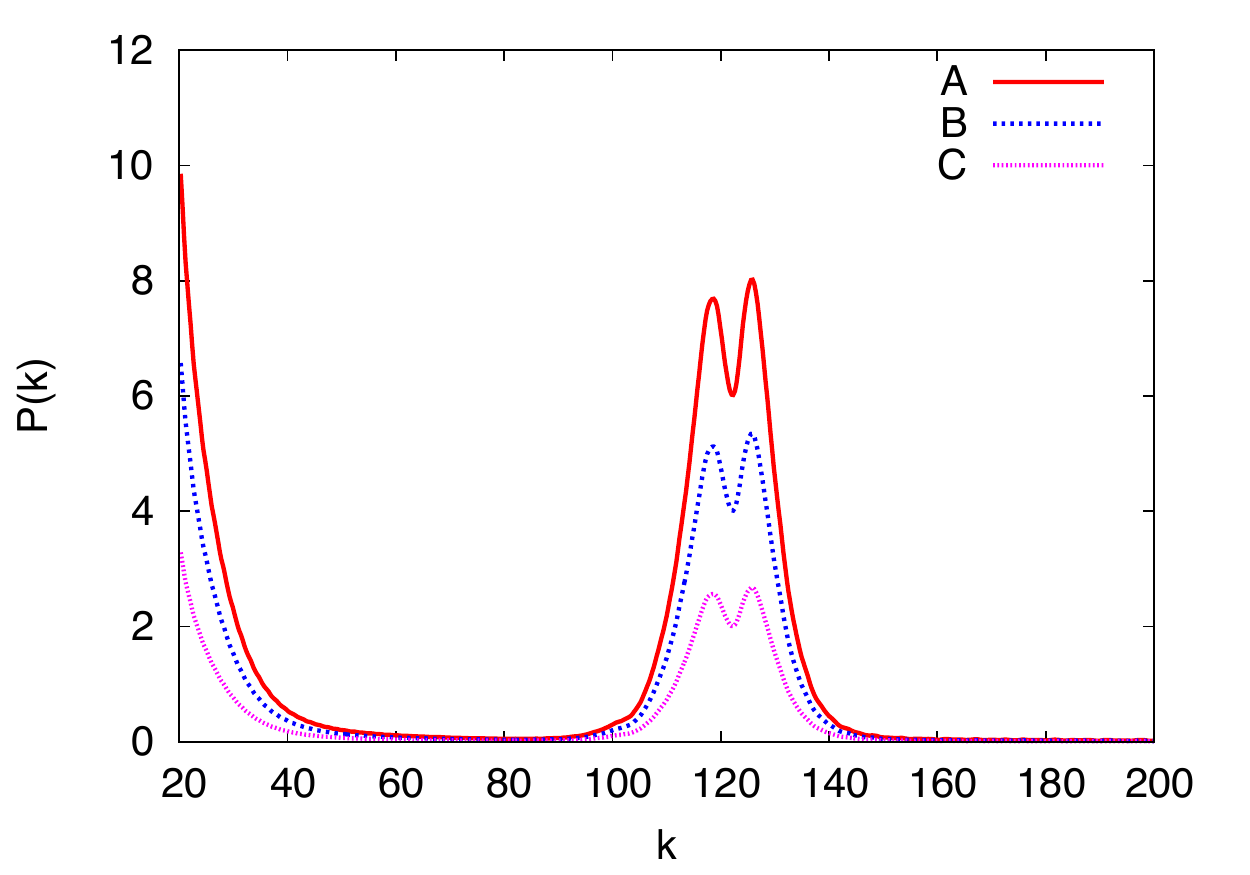}
\\
\includegraphics[scale=1,width=8cm]{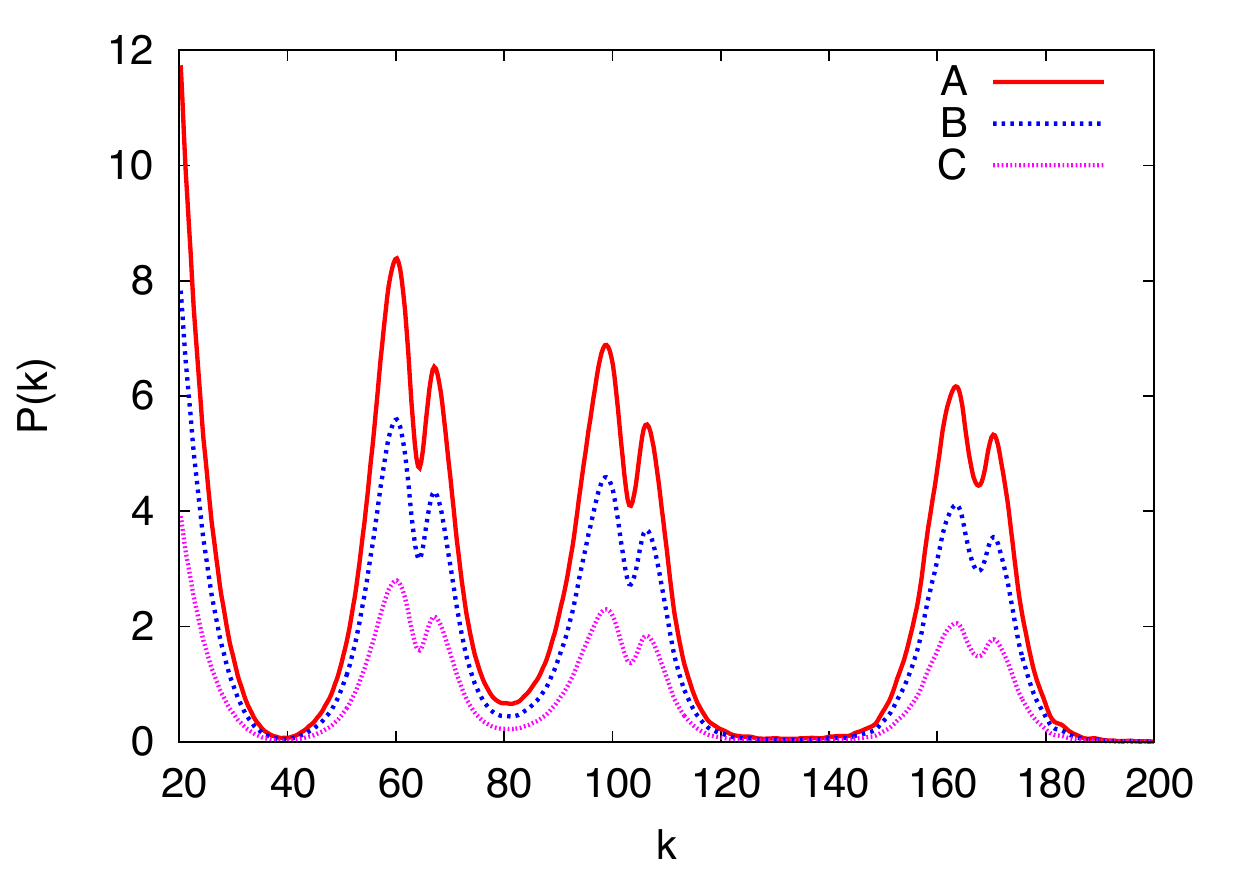}
\includegraphics[scale=1,width=8cm]{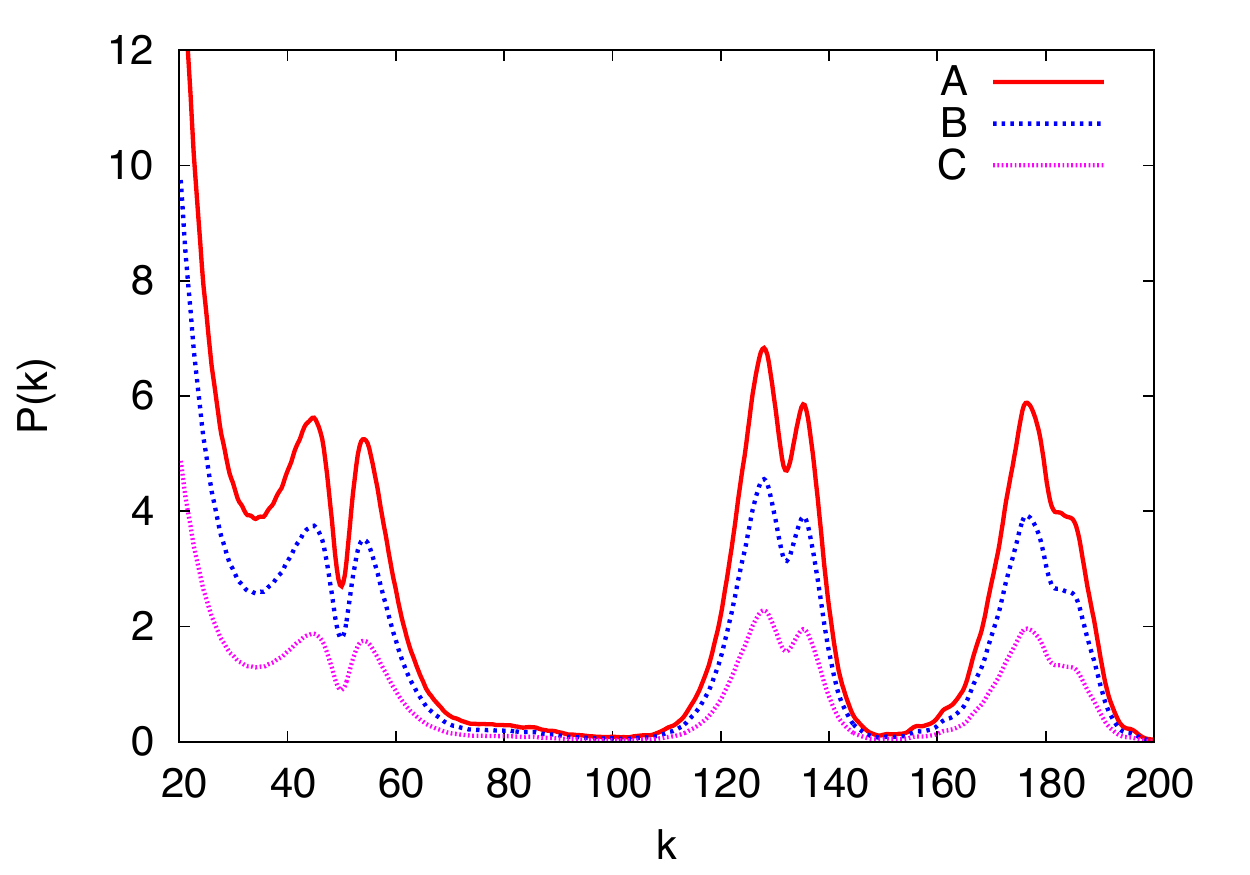}
\end{tabular}
\end{center}
\caption{ Power spectrum versus k for $L=5000$ km(Upper-Left), $L=8000$ km(Upper-Right),
$L=11000$ km(Lower-Left), $L=12000$ km(Lower-Right) with 
$\Delta_\nu/E_\nu=0.03/\sqrt{E_\nu/\textrm{MeV}}+0.5 \times E_\nu/m_p$.  
Earth matter effect calculated numerically using PREM density profile with 5 layers.
Line A:   $N=3.\times 10^4$ ;
Line B:   $N=2.\times 10^4$ ;
Line C:   $N=1.\times 10^4$ .}
\label{figure3}
\end{figure}

\begin{figure}[tb]
\begin{center}
\begin{tabular}{cc}
\includegraphics[scale=1,width=8cm]{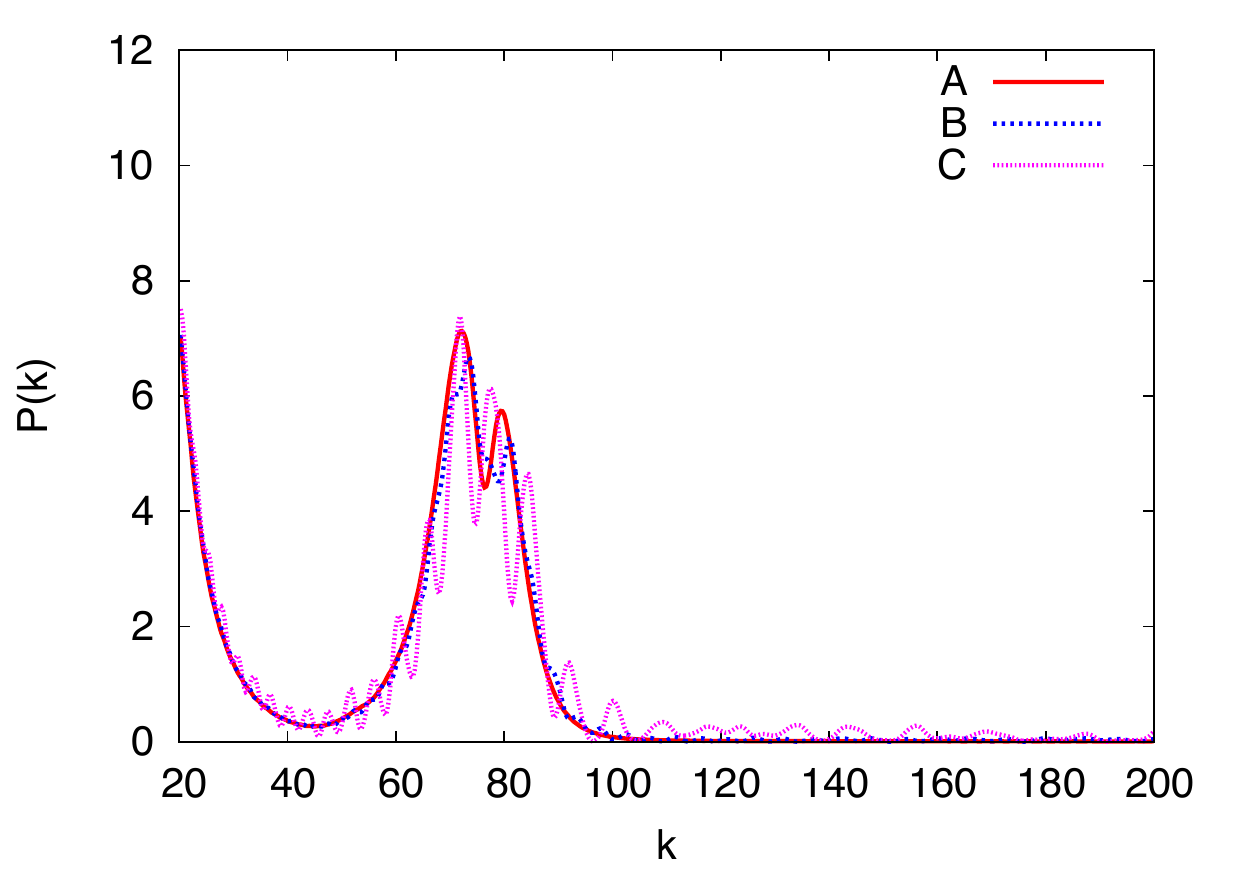}
\includegraphics[scale=1,width=8cm]{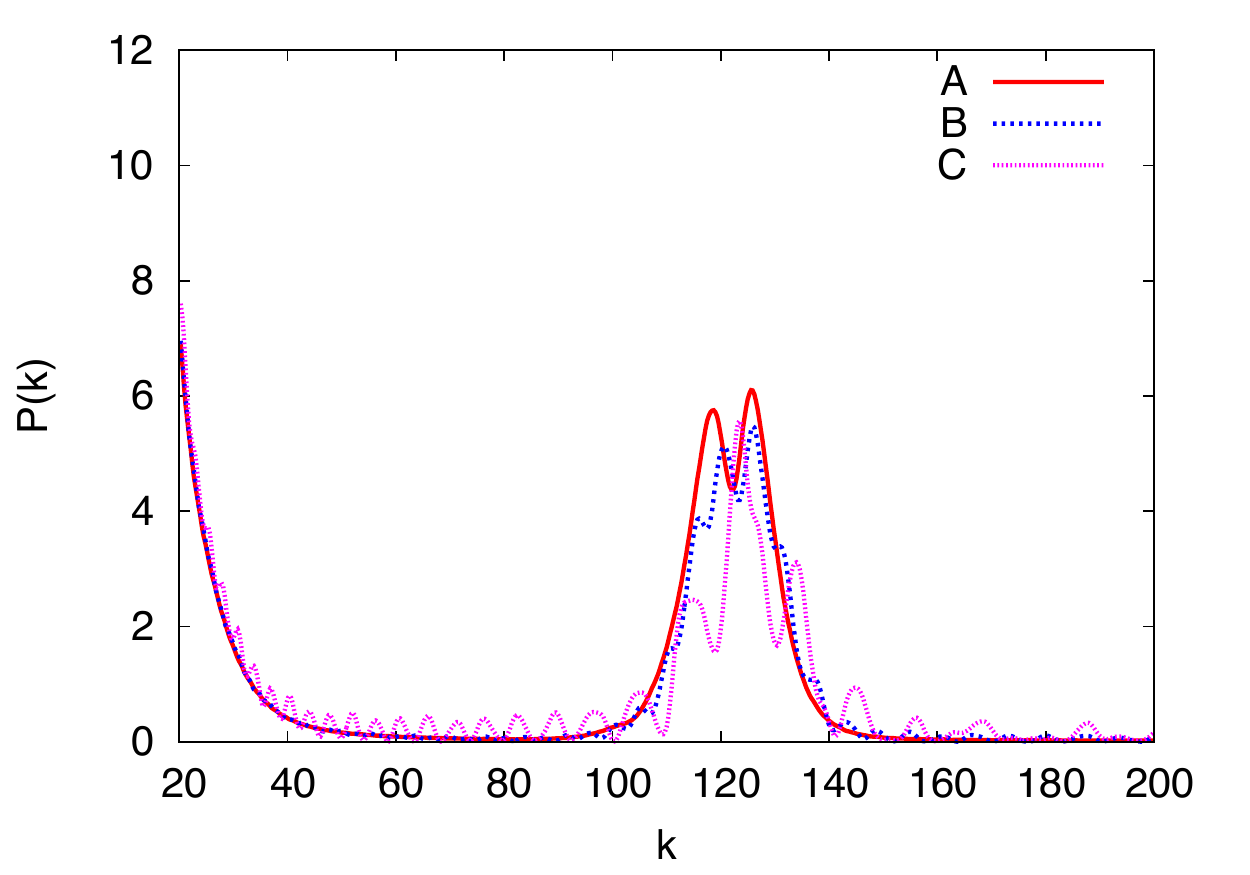}
\\
\includegraphics[scale=1,width=8cm]{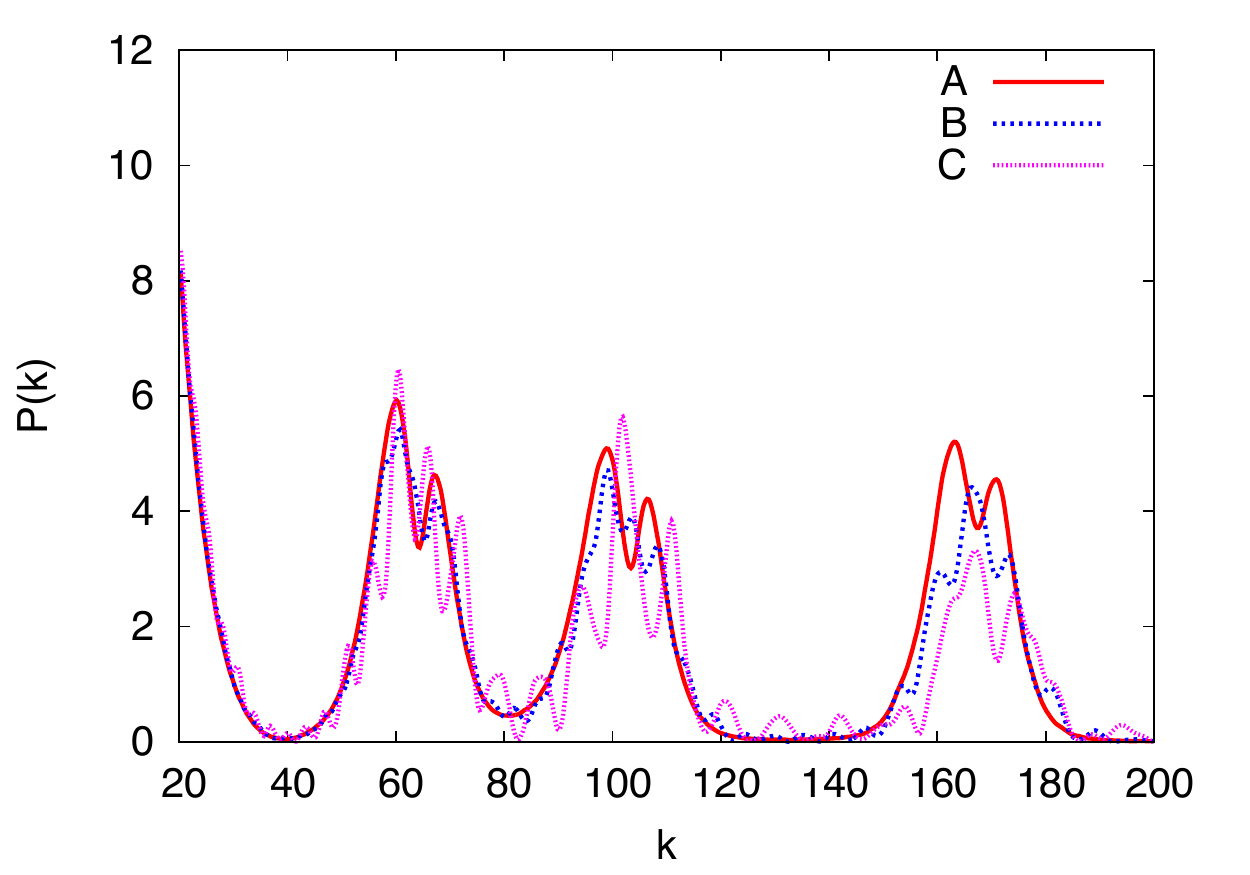}
\includegraphics[scale=1,width=8cm]{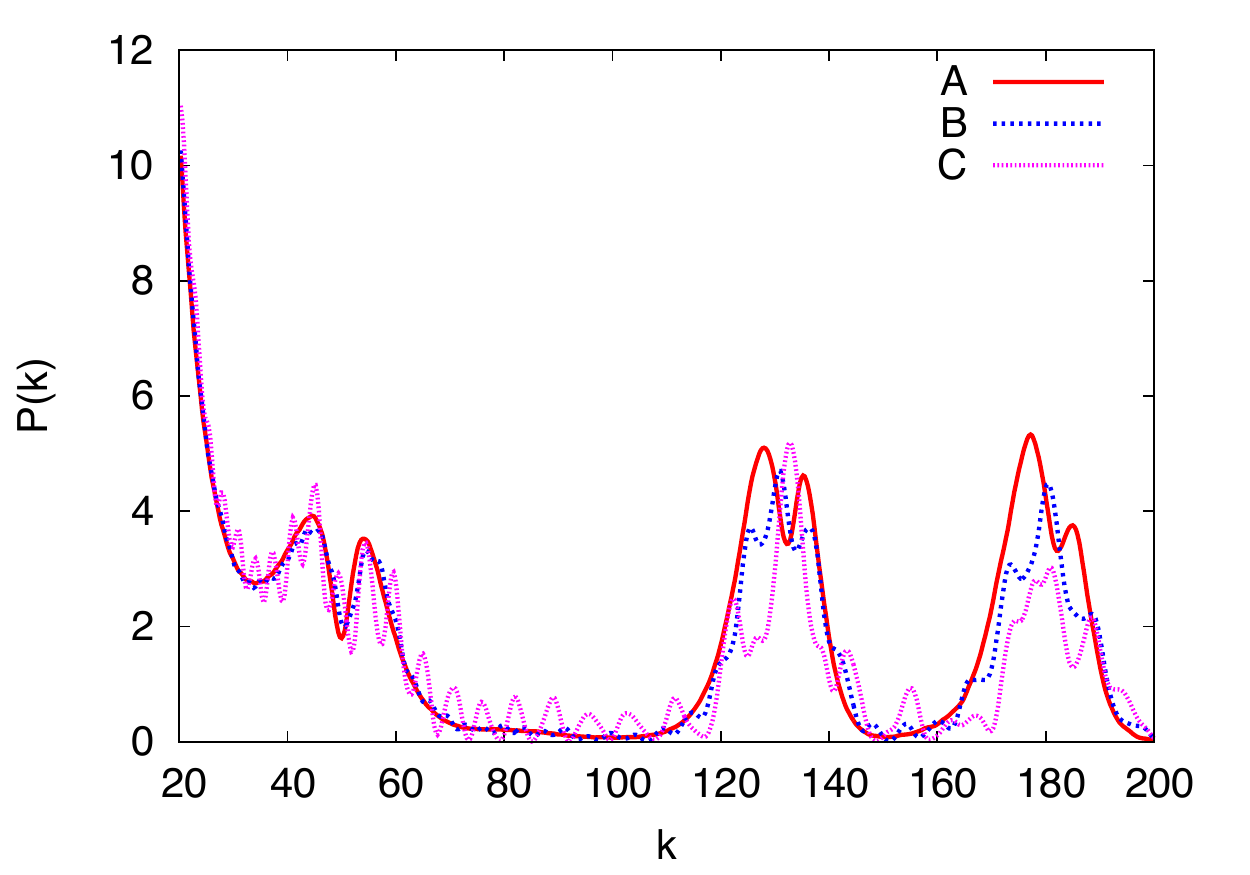}
\end{tabular}
\end{center}
\caption{ Power spectrum versus k for $L=5000$ km(Upper-Left), $L=8000$ km(Upper-Right),
$L=11000$ km(Lower-Left), $L=12000$ km(Lower-Right) and for $N=2.\times 10^4$ .
Line A:  $\Delta_\nu/E_\nu=0.03/\sqrt{E_\nu/\textrm{MeV}}+0.2 \times E_\nu/m_p$ ;
Line B:  $\Delta_\nu/E_\nu=0.10/\sqrt{E_\nu/\textrm{MeV}}+0.2 \times E_\nu/m_p$ ;
Line C:  $\Delta_\nu/E_\nu=0.20/\sqrt{E_\nu/\textrm{MeV}}+0.2 \times E_\nu/m_p$ .
Earth matter effect calculated numerically using PREM density profile with 5 layers.
 }
\label{figure4}
\end{figure}

(\ref{reg1a}) and (\ref{reg1b}) tell us something important about the Earth matter effect in oscillation of SNe neutrinos.
Major $\Delta V_i$ contributing to the regeneration factor are density jumps at the surface and the mantle-core crossing boundaries of the Earth.
Consequently, for neutrinos not crossing the core($L< 10690$km),  (\ref{reg1a}) and (\ref{reg1b}) are
 dominated by the term associated with the density jump when neutrinos entering the Earth.
In this case,  oscillation of SNe neutrino in the Earth is basically an one frequency oscillation(one distance from the density jump to
detector) with a magnitude increasing with energy,
as can be seen for the case $L=8000$km in Fig. (\ref{figure1}).  For core-crossing neutrinos, the regeneration factors get two more
major contributions from crossing core-mantle boundaries. In this case, 
 (\ref{reg1a}) and (\ref{reg1b}) tell us that oscillation of SNe neutrino in the Earth is basically an oscillation with three frequencies, 
 i.e. with three oscillation phases in three terms in (\ref{reg1a}) and (\ref{reg1b}).
These oscillation patterns can be figured out by Fourier transforming the event rate of SNe neutrinos detected in detector.

Introducing~\cite{SNeNeu1}
\bea
G(k)= \frac{1}{\sqrt{N}}\int dy ~F(y) e^{ik y}, \label{Fourier1} 
\eea
where $y=12.5$MeV$/E$,  k a number denoting the modes, $F(y)$ the event rate and $N$ the total number of events
which is introduced for purpose of normalization. Earth matter effect in oscillation of neutrinos would give a modulation of
neutrino spectrum and give rise to peaks in the power spectrum
\bea
P(k)=|G(k)|^2
\eea
in the plot of $P(k)$ versus $k$. For practical use, (\ref{Fourier1}) can be re-written as
\bea
G(k)=\frac{1}{\sqrt{N}} \sum_{\textrm{energy bin i}}  \frac{N_i}{\Delta_{y_i}} \int_{\Delta_{y_i}} ~dy_i~e^{i k y_i} \label{Fourier2}
\eea
where $N_i$ is the number of events in i$^{th}$ energy bin, $\Delta_{y_i}$ the width of y in i$^{th}$ energy bin.
In (\ref{Fourier2}), an average over phase in each energy bin is introduced.  As we said before, the energy bin
chosen in (\ref{Fourier2}) can be larger than that defined by the energy resolution. We can actually vary the
width of energy bin in analysis as long as it's allowed by the energy resolution of a specific detector.
The significance of $P(k)$ is that if 1$\sigma $ fluctuation is introduced in number of events in each energy bin,
$P(k)$ is expected to be around 1. If  peaks with $P(k)$ much larger than one are observed in the
power spectrum, they are possible signals of Earth matter effect in neutrino oscillation~\cite{SNeNeu1}.

We assume NH of neutrino masses in the
remaining part of this article. For IH, the Earth matter effect is expected to appear 
in neutrino sector of SNe neutrinos, not in anti-neutrino sector, as can be seen in (\ref{reg0a}) and (\ref{reg0b}).
Analysis of Earth matter effect in neutrino sector of SNe neutrinos, e.g. in events of $\nu_e$,  can
be done for a detector such as LaTPC which is most sensitive
to $\nu_e$ events,  in a way similar to what shown below for ${\bar \nu}_e$ events.

\begin{figure}[tb]
\begin{center}
\begin{tabular}{cc}
\includegraphics[scale=1,width=8cm]{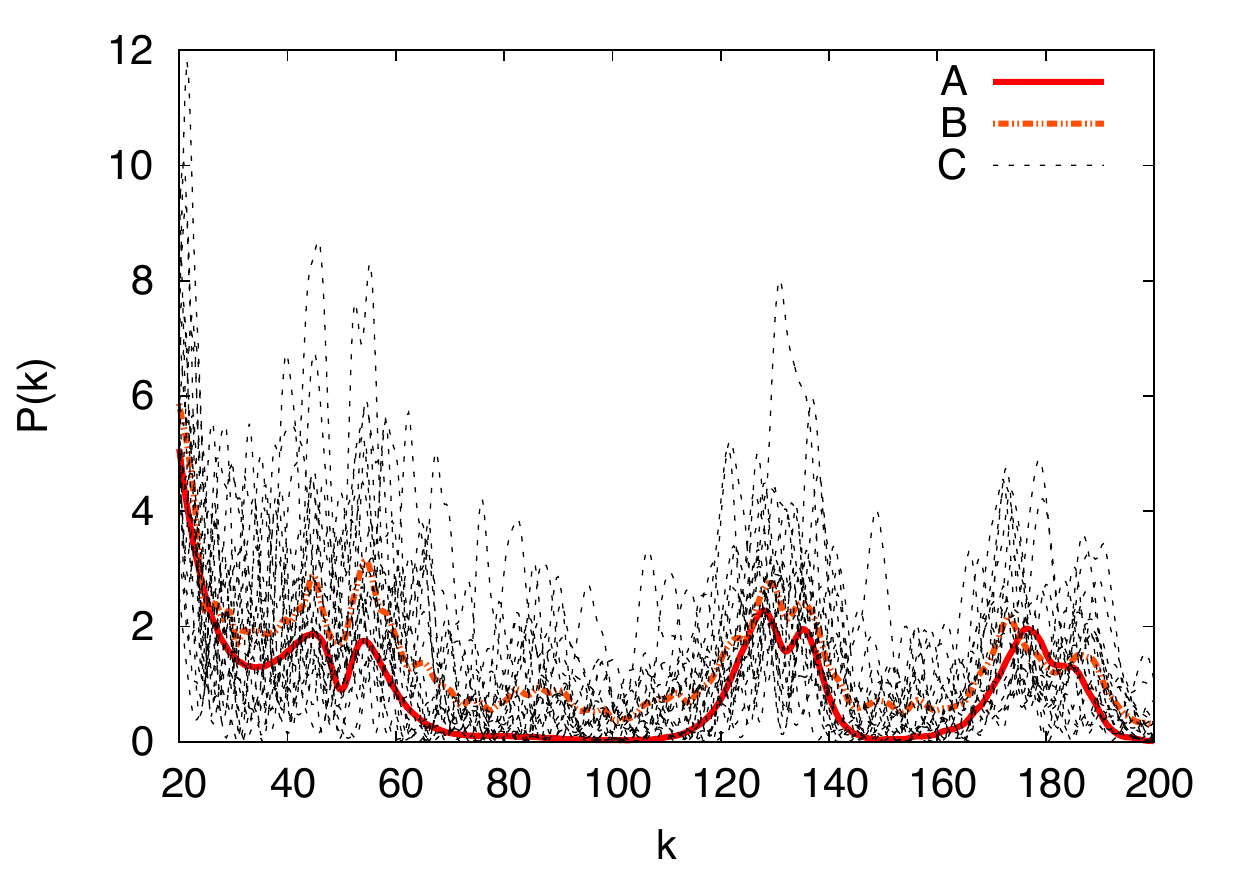}
\includegraphics[scale=1,width=8cm]{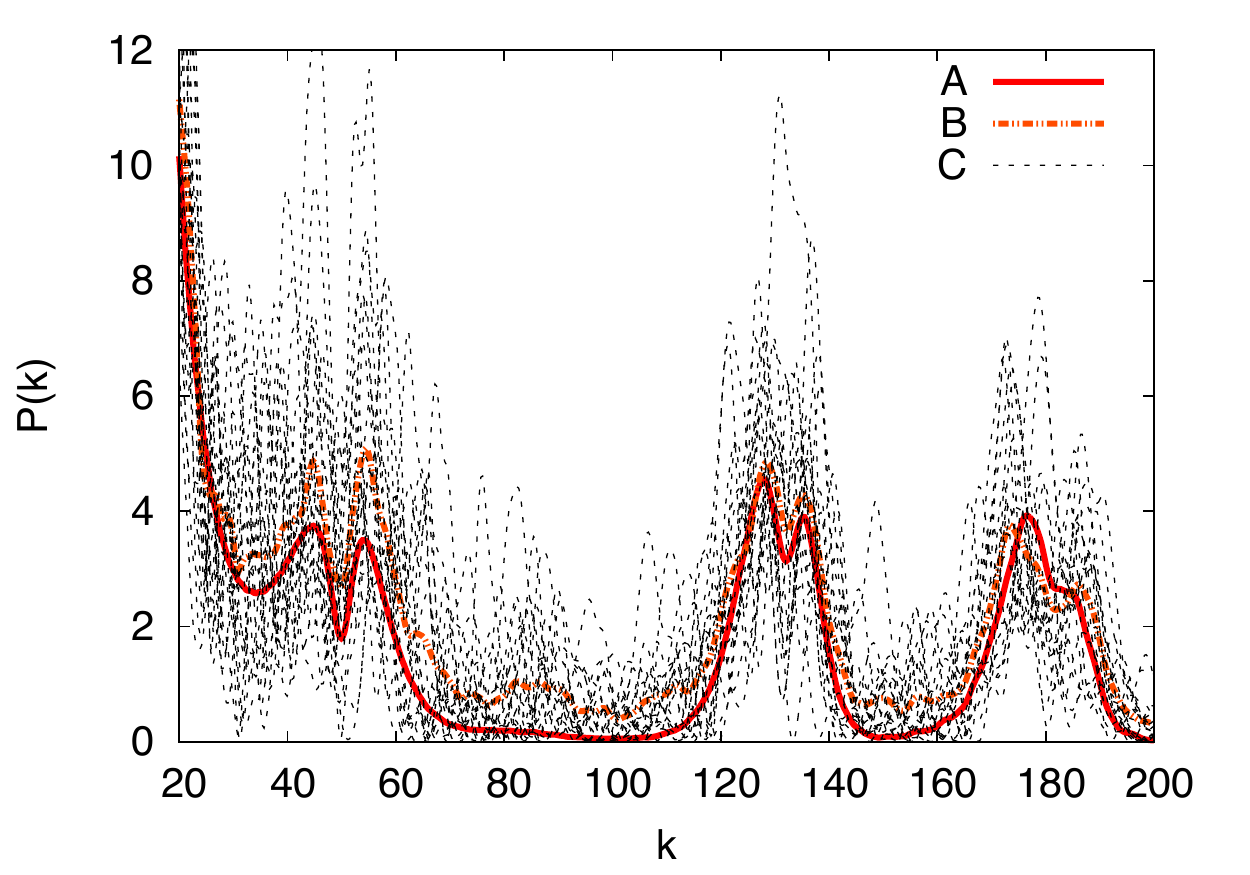}
\\
\includegraphics[scale=1,width=8cm]{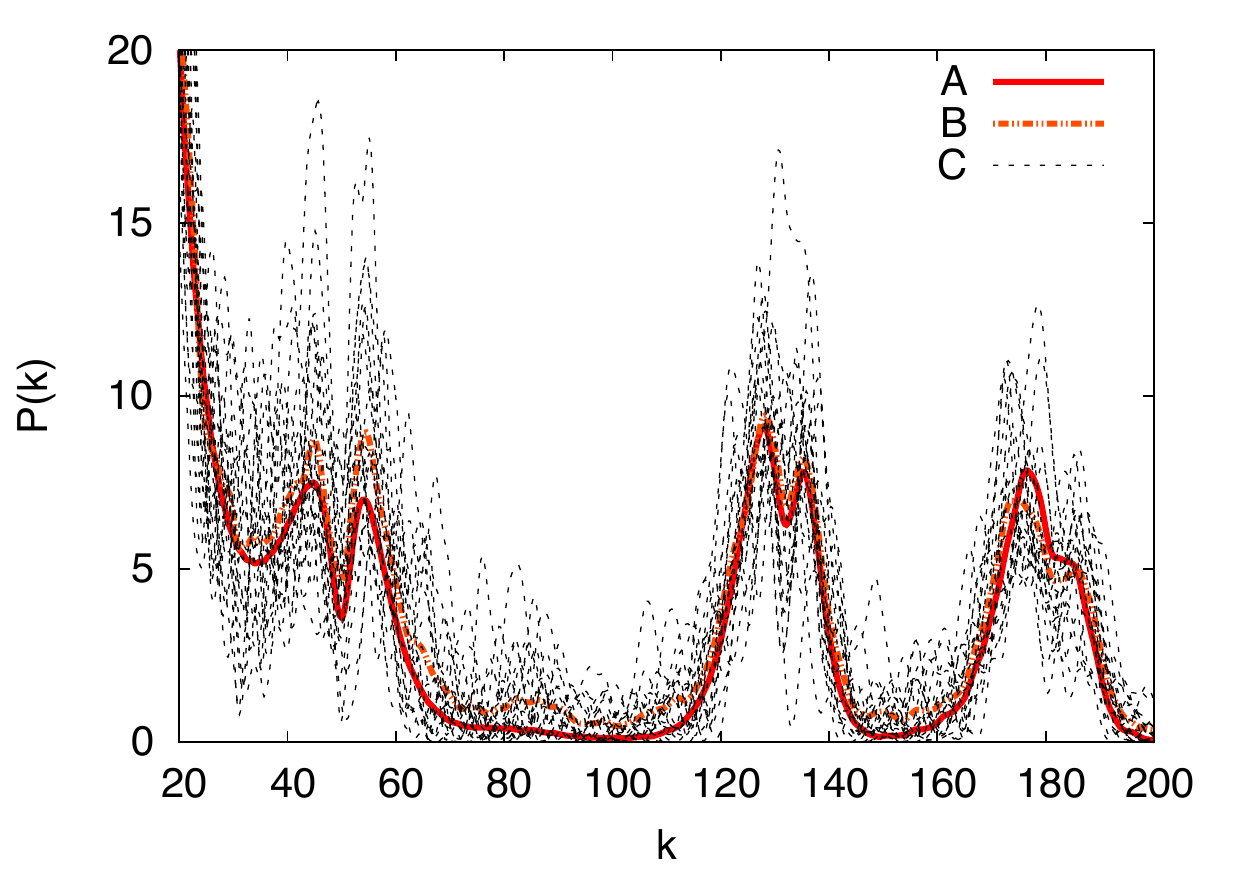}
\includegraphics[scale=1,width=8cm]{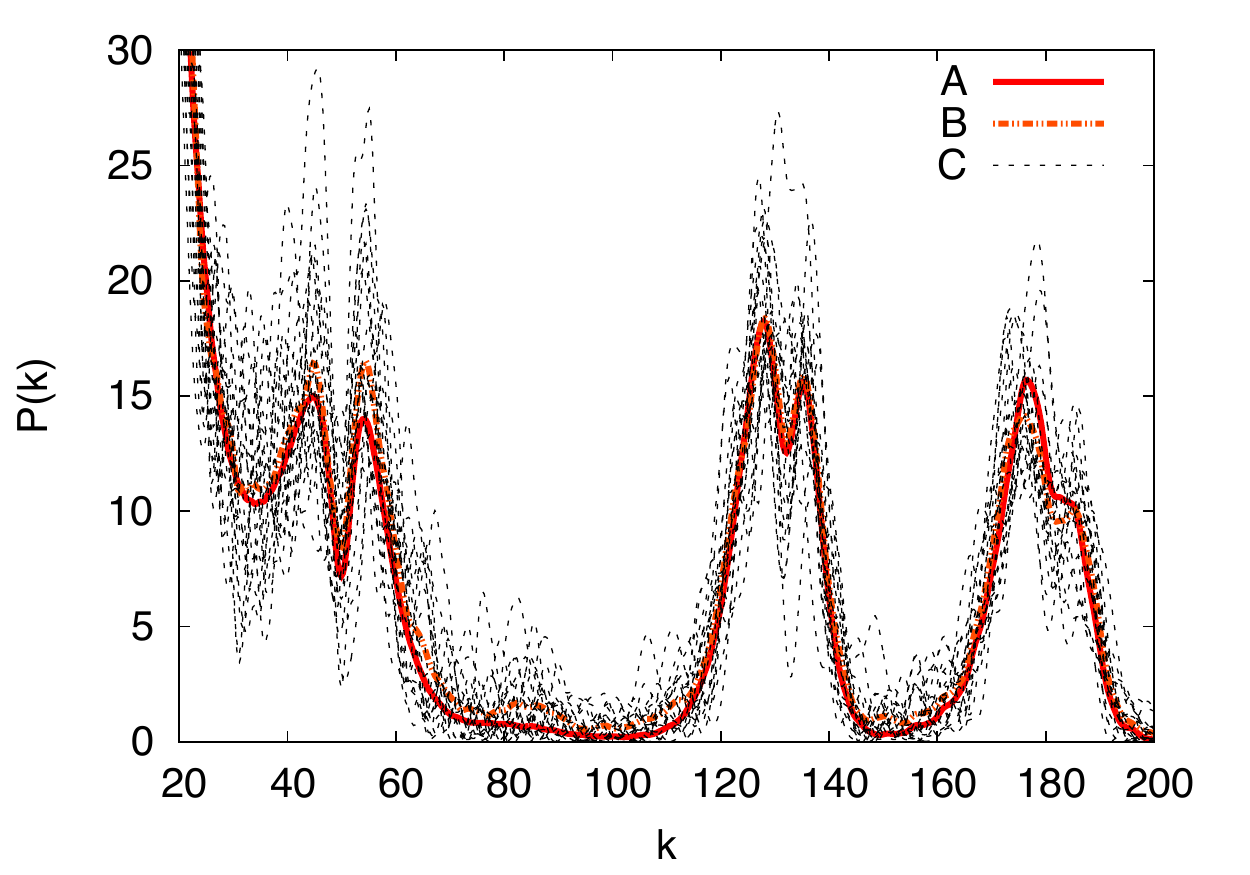}
\end{tabular}
\end{center}
\caption{ Power spectrum versus k for  $N=1.\times 10^4$(Upper-Left),  $N=2.\times 10^4$(Upper-Right),
 $N=4.\times 10^4$(Lower-Left),  $N=8.\times 10^4$(Lower-Right) , with
$L=12000$ km and $\Delta_\nu/E_\nu=0.03/\sqrt{E_\nu/\textrm{MeV}}+0.5 \times E_\nu/m_p$.
Line A:  theoretical expectation;
Line B:  average over 20 samples;
Line C:  20 samples with $1\sigma$ fluctuation.
Earth matter effect calculated numerically using PREM density profile with 5 layers. }
\label{figure5}
\end{figure}

In Fig. \ref{figure2} we plot the power spectrum $P(k)$ versus k for IBD events of ${\bar \nu}_e$ 
without fluctuation and for various baselines. 
When making the plots,  we use the fitted primary spectrum averaged over time~\cite{spectrum}
\bea
F^0_{\bar \nu}(E_\nu)=\frac{\Phi_{\bar \nu}}{\langle E_\nu \rangle} \frac{(1+\alpha)^{1+\alpha}}{\Gamma(1+\alpha)}
\bigg(  \frac{E_\nu}{\langle E_\nu \rangle}\bigg)^\alpha e^{-(1+\alpha) E_\nu/\langle E_\nu \rangle },
\label{fittedSpec}
\eea 
$\Phi_\nu$ represents the flux integrated over $E_\nu$.  For simplicity, we have neglected possible modification
of collective effect in neutrino oscillation to the neutrino primary spectrum since recent
studies show that this effect seems to be suppressed under some circumstances~\cite{suppression}, in particular
for accretion phase and earlier cooling phase.
We expect that including collective effect in neutrino oscillation would refine our analysis on the
effect of energy resolution and angular resolution of detectors but would not change the main conclusion 
about their effect presented in this article.

In Fig. \ref{figure2}, we take $\langle E_\nu \rangle= 11$ MeV and $\alpha=3$ for $F^0_{{\bar \nu}_e}$,
$\langle E_\nu \rangle= 18$ MeV and $\alpha=2$ for $F^0_{{\bar \nu}_X}$,  and
$\Phi_{{\bar \nu}_e}=2 \Phi_{{\bar \nu}_X}$.   $\Phi_{{\bar \nu}_e}$ and $\Phi_{{\bar \nu}_X}$ describe
the absolute intensities of neutrinos arriving at detector. The actual values of $\Phi_{{\bar \nu}_e}$ and
$\Phi_{{\bar \nu}_X}$ depend on the
distance of SN from the Earth and we do not fix. These parameters represents some properties 
of SNe neutrinos during the accretion phase.  In all figures to be shown later, we always use these parameters and
(\ref{fittedSpec}) for the initial spectrum of SNe neutrinos. In all these plots, 
we always compute the power spectrum for SNe neutrinos in energy range from 3 MeV to 70 MeV. 
We compute Earth matter effect in neutrino oscillation and then calculate the spectrum
of SNe neutrinos and the event rate arriving at detector.  For  each energy bin, we average the regeneration factor in each energy bin and 
obtain event rate of neutrinos in each energy bin.

\begin{figure}[tb]
\begin{center}
\begin{tabular}{cc}
\includegraphics[scale=1,width=8cm]{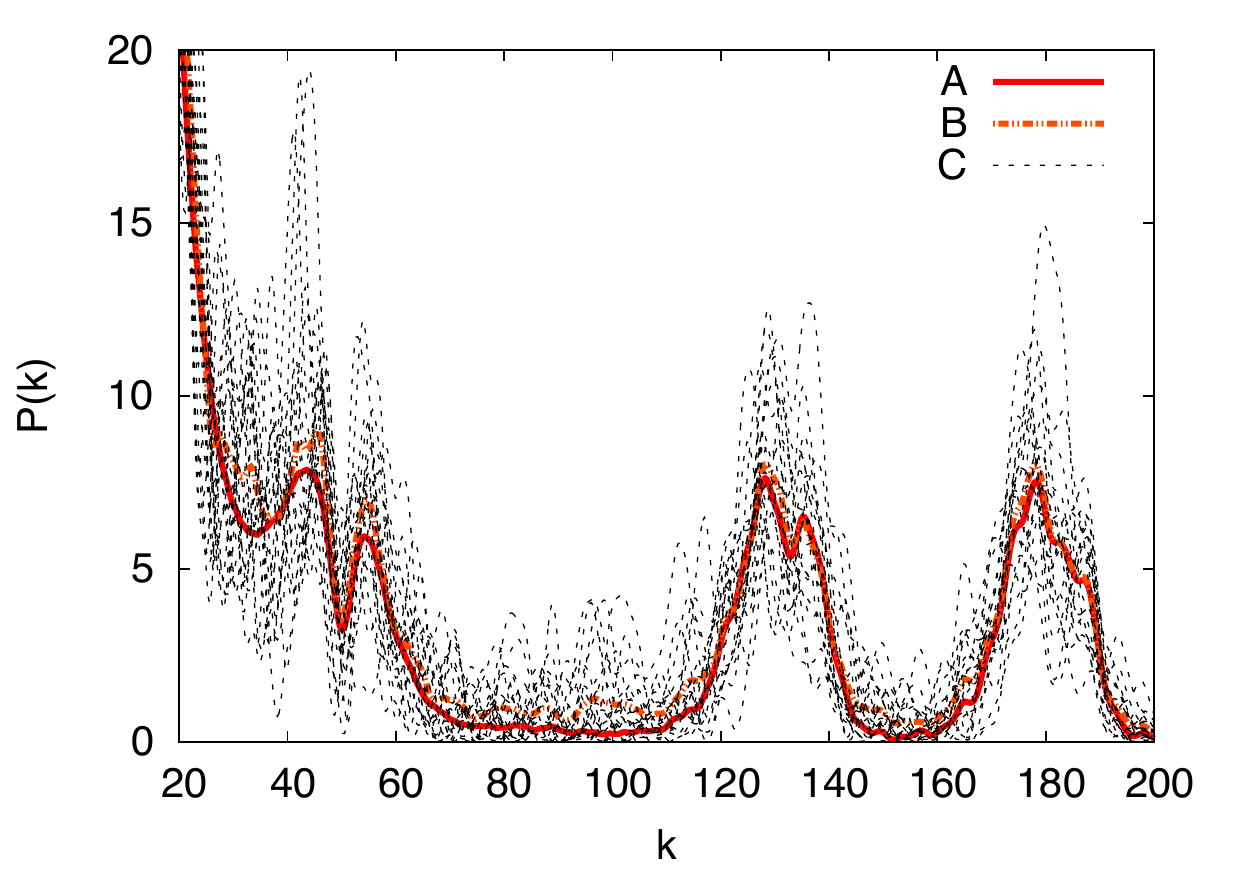}
\includegraphics[scale=1,width=8cm]{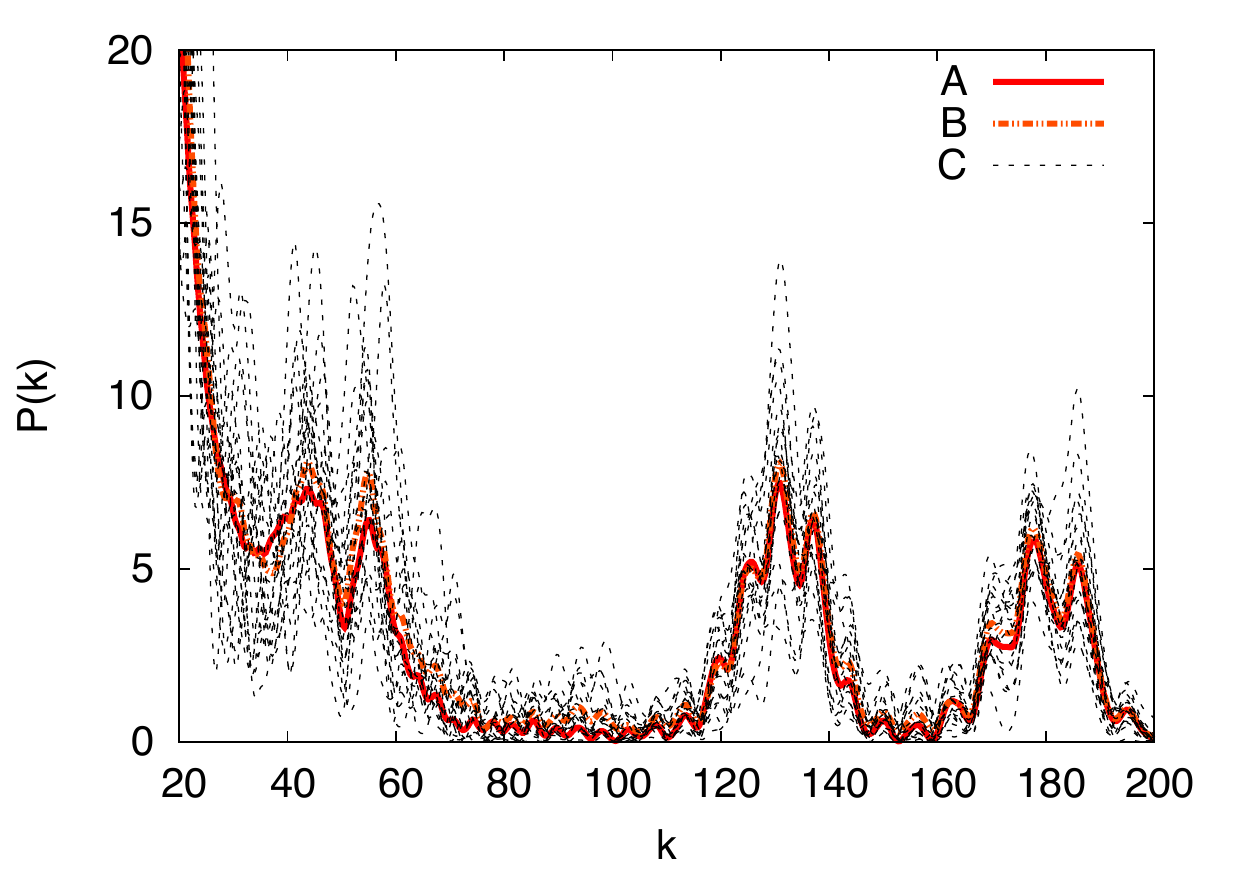}
\\
\includegraphics[scale=1,width=8cm]{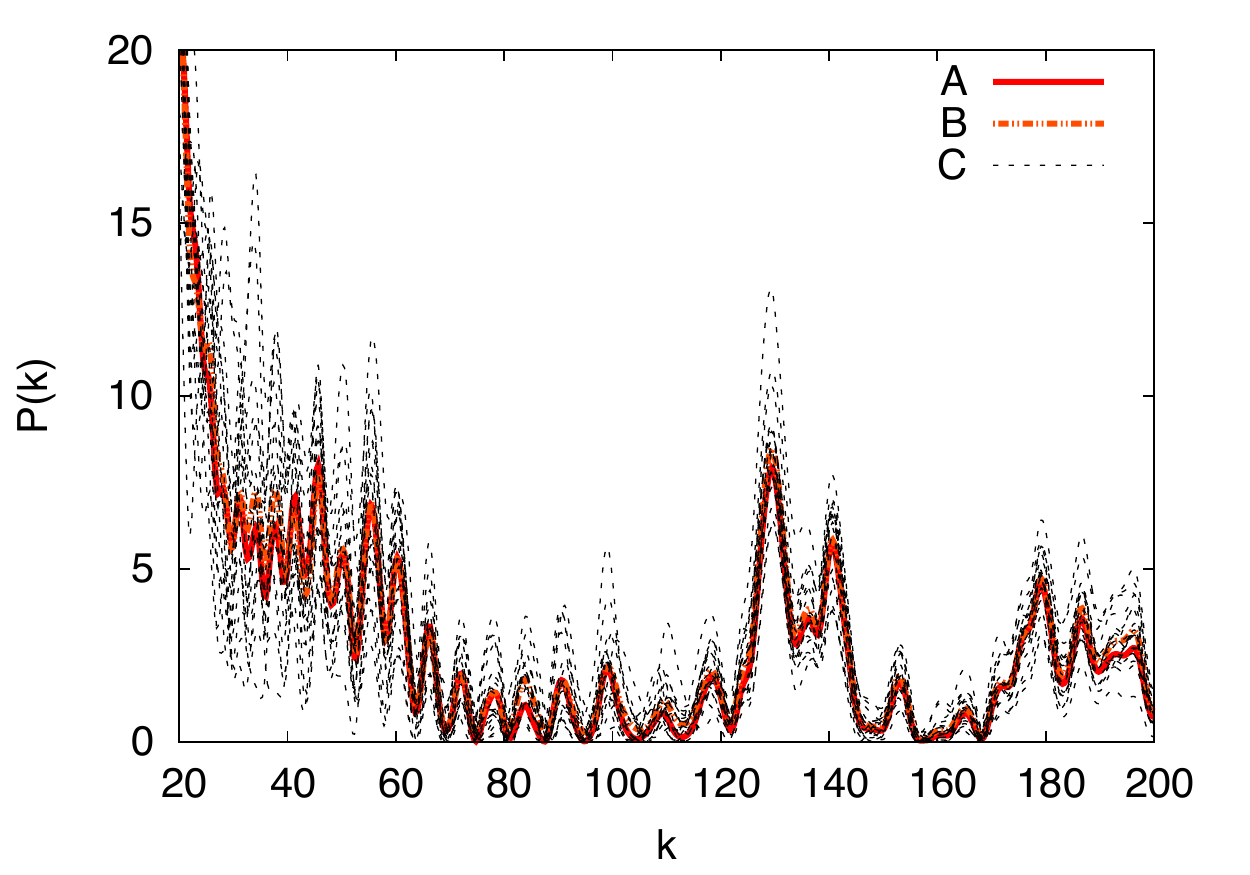}
\includegraphics[scale=1,width=8cm]{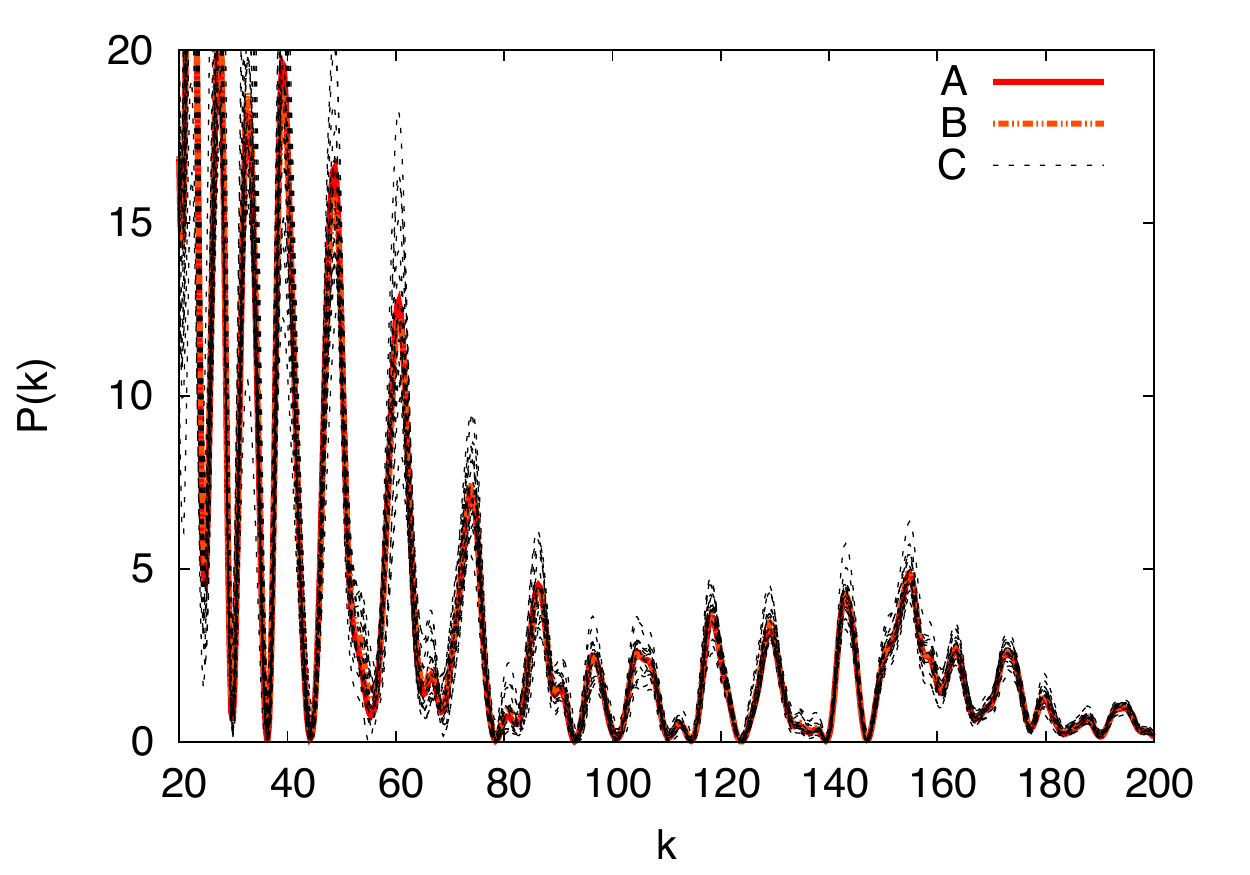}
\end{tabular}
\end{center}
\caption{ Power spectrum versus k for $L=12000$ km and  $N=4.\times 10^4$.
Upper-Left: $\Delta_\nu/E_\nu=0.05/\sqrt{E_\nu/\textrm{MeV}}+0.5 \times E_\nu/m_p$;
Upper-Right: $\Delta_\nu/E_\nu=0.10/\sqrt{E_\nu/\textrm{MeV}}+0.5 \times E_\nu/m_p$;
Lower-Left: $\Delta_\nu/E_\nu=0.20/\sqrt{E_\nu/\textrm{MeV}}+0.5 \times E_\nu/m_p$;
Lower-Right: $\Delta_\nu/E_\nu=0.50/\sqrt{E_\nu/\textrm{MeV}}+0.5 \times E_\nu/m_p$.
Line A:  theoretical expectation;
Line B:  average over 20 samples;
Line C:  20 samples with $1\sigma$ fluctuation.
Earth matter effect calculated numerically using PREM density profile with 5 layers. }
\label{figure6}
\end{figure}

In Fig. \ref{figure2}, one can see that there are three peaks for core-crossing
baselines and only one peak for baselines not crossing the core of the Earth, in agreement with discussions
given above for Fig. \ref{figure1}. That is, for core-crossing neutrinos, three peaks are coming from contributions of two core-mantle density jumps
and one mantle-surface density jump when neutrinos entering the Earth. For baseline not crossing the core,
there is no core-mantle density jump and there is only one peak.
Peaks with larger k values correspond to oscillations
with longer distance from the density jump to detector. The shorter the distance from the density jump, 
the smaller the value of k of the position of peak, as can be seen
in plots of $L=5000$ km and $L=8000$ km respectively. 
This can be understood by approximating each term in (\ref{reg1a}) and (\ref{reg1b})
by an average of density of matter so that phase $\Phi_i$ and ${\bar \Phi}_i$ can be written as 
$\Phi_i=\frac{\Delta_i}{4E} L_i$ and $\Phi_i=\frac{{\bar \Delta}_i}{4E} L_i$ where
$L_i$ is the distance from the i$^{th}$ density jump to detector. $\Delta_i$ and ${\bar \Delta}_i$  are the effective
mass squared differences of  neutrino and anti-neutrino in matter respectively, and for energy of SNe neutrinos
and Earth matter density one can take $\Delta_i \approx {\bar \Delta}_i$ as a first order approximation.
So contribution from density
jump farther away from the detector, i.e. for larger $L$,  corresponds to larger k value in Fourier mode in (\ref{Fourier1}):  
$k$ value of peak $\propto L$.
As a further note, one can also approximate the matter density as constant in the mantle and in the core
of Earth separately. In this approximation, (\ref{reg1a}) and (\ref{reg1b}) can be reduced to
formula presented in ~\cite{SNeNeu1,SNeNeu2} and one can similarly conclude that larger k value of peak corresponds
to longer distance from the density jump to detector~\cite{SNeNeu2}.

One can see in Fig. \ref{figure2} that the lines with oscillation clearly distinguish with the line with no oscillation.
The higher  the peak, the more
significant the signal of the Earth matter effect.   
In the plots, we have shown effects of the size of energy bin on the signal strength. One can see that
increasing the width of energy bin reduces the signal strength.  Increasing the width of energy bin reduces the
signal strength more significantly for larger k value which corresponds to longer baseline of oscillation. 
This is because for longer baseline of oscillation  increasing the energy bin introduces more averaging over oscillation phase.
Hence, as expected, effect of changing energy bin is more significant for peak with larger k value.

In Fig. \ref{figure3},  we compare different cases with number of events changed. One can see that the signal
strength increases as the number of events increases. In Fig. \ref{figure4}, we compare effects of different values of $r_e$
on the signal strength. For larger $r_e$, i.e. with larger width of energy bin, the signal strength is also reduced, similar to the case in Fig. \ref{figure2}.
The effect of changing $r_e$ is also 
more significant for peaks with larger values of k, which correspond to contributions of density jumps more far away from
detector. A difference between changing $r_e$ and changing $r_a$, as can be seen in Fig. \ref{figure2} and Fig. \ref{figure4},
is that changing $r_a$ has more significant impact on the signal strength. This is because changing $r_a$ affects
more significantly energy bins at high energy part while changing $r_e$ affects more significantly energy bins at low energy part.
Since the Earth matter effect in oscillation of SNe neutrino shows up mainly in high energy part of the spectrum, as can be seen in Fig. \ref{figure1},
the signal strength is more sensitive to changing $r_a$ than changing $r_e$. These discussions
show us that a reasonable angular resolution, e.g. with $r_a\lsim 0.5$, is very helpful to get a good signal strength of
Earth matter effect in power spectrum $P(k)$.  

\begin{figure}[tb]
\begin{center}
\begin{tabular}{cc}
\includegraphics[scale=1,width=8cm]{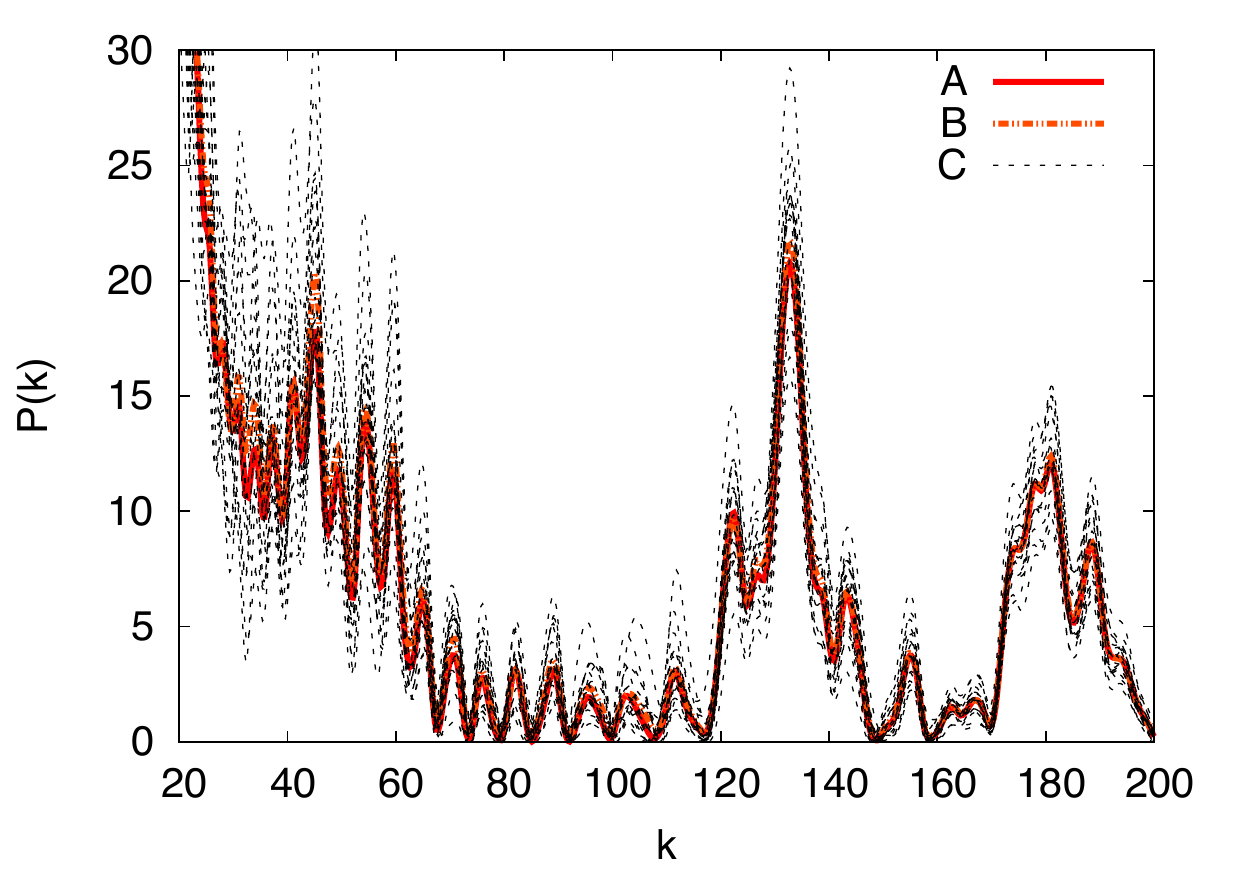}
\includegraphics[scale=1,width=8cm]{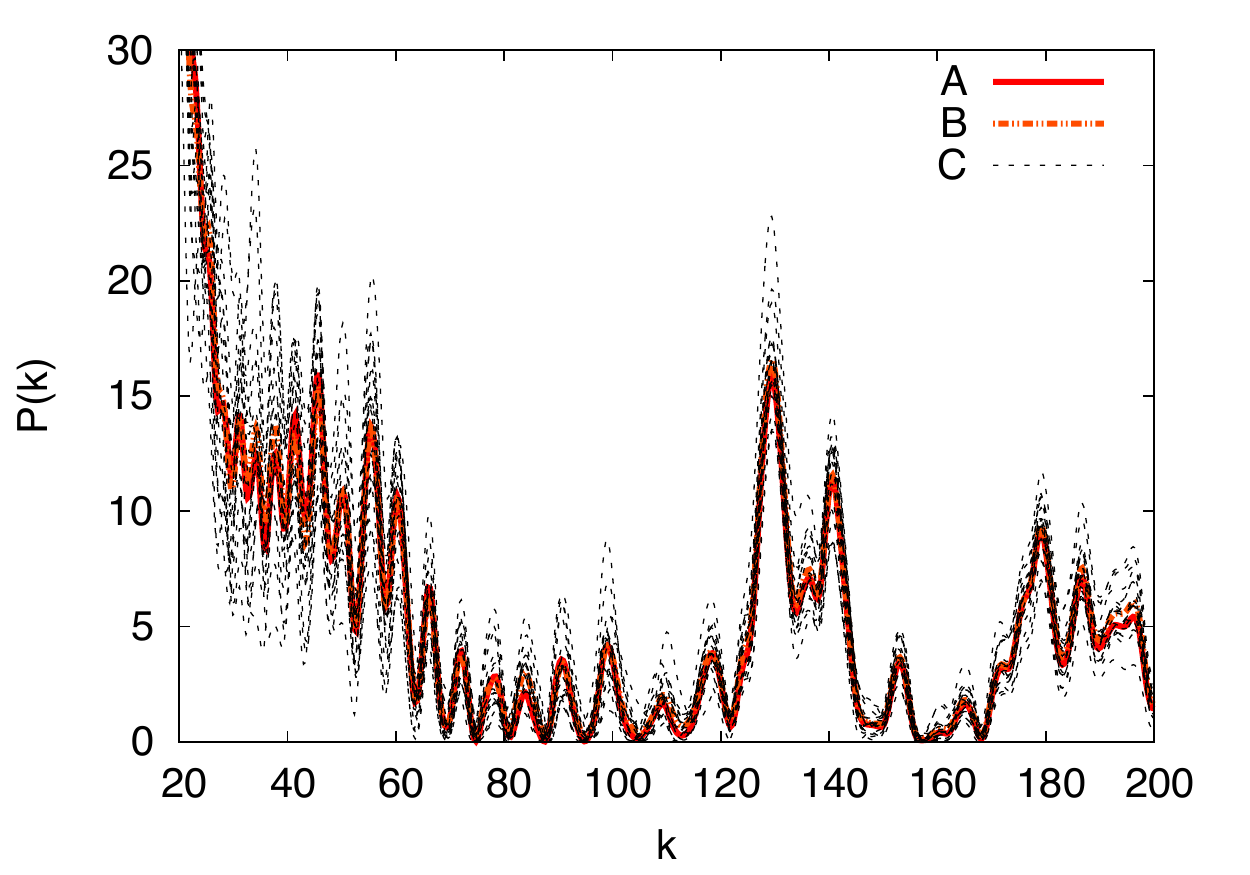}
\\
\includegraphics[scale=1,width=8cm]{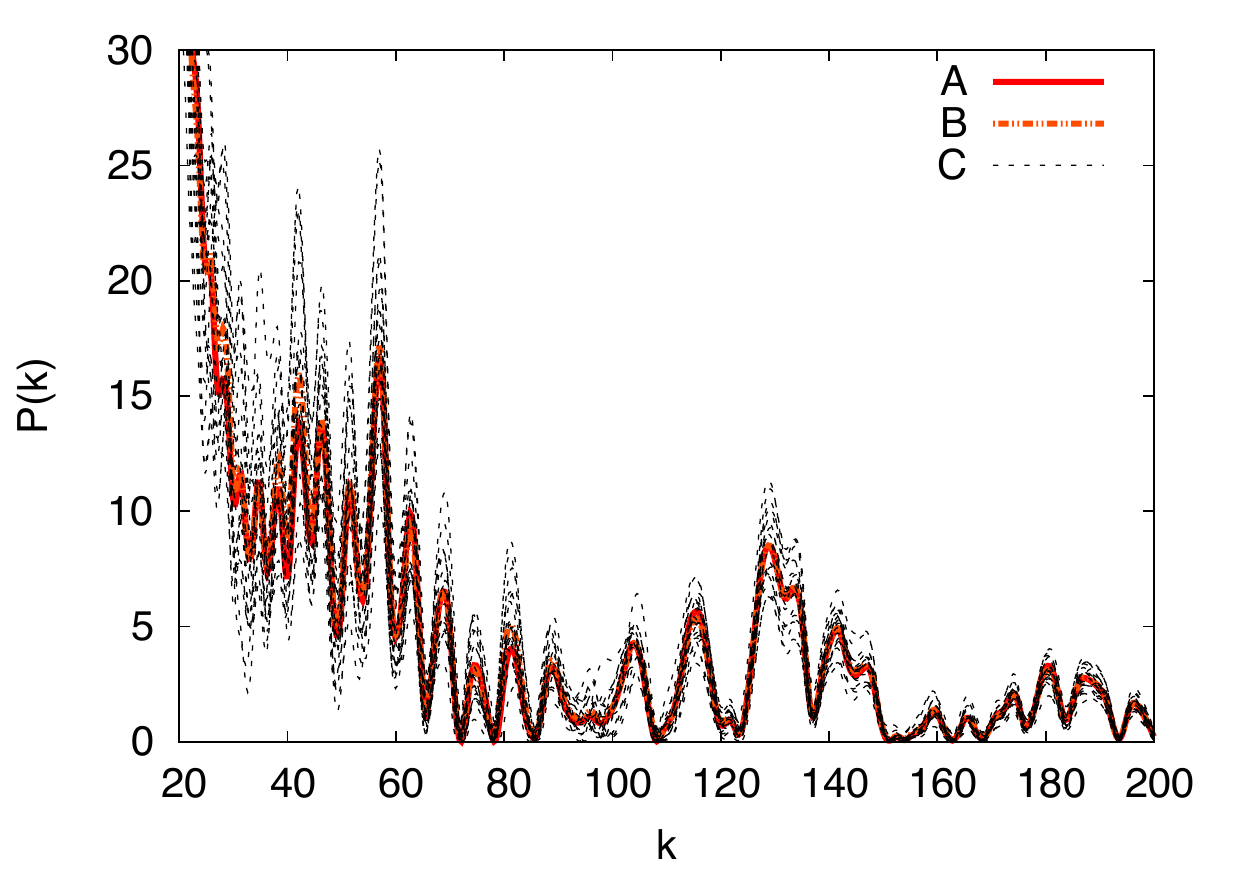}
\includegraphics[scale=1,width=8cm]{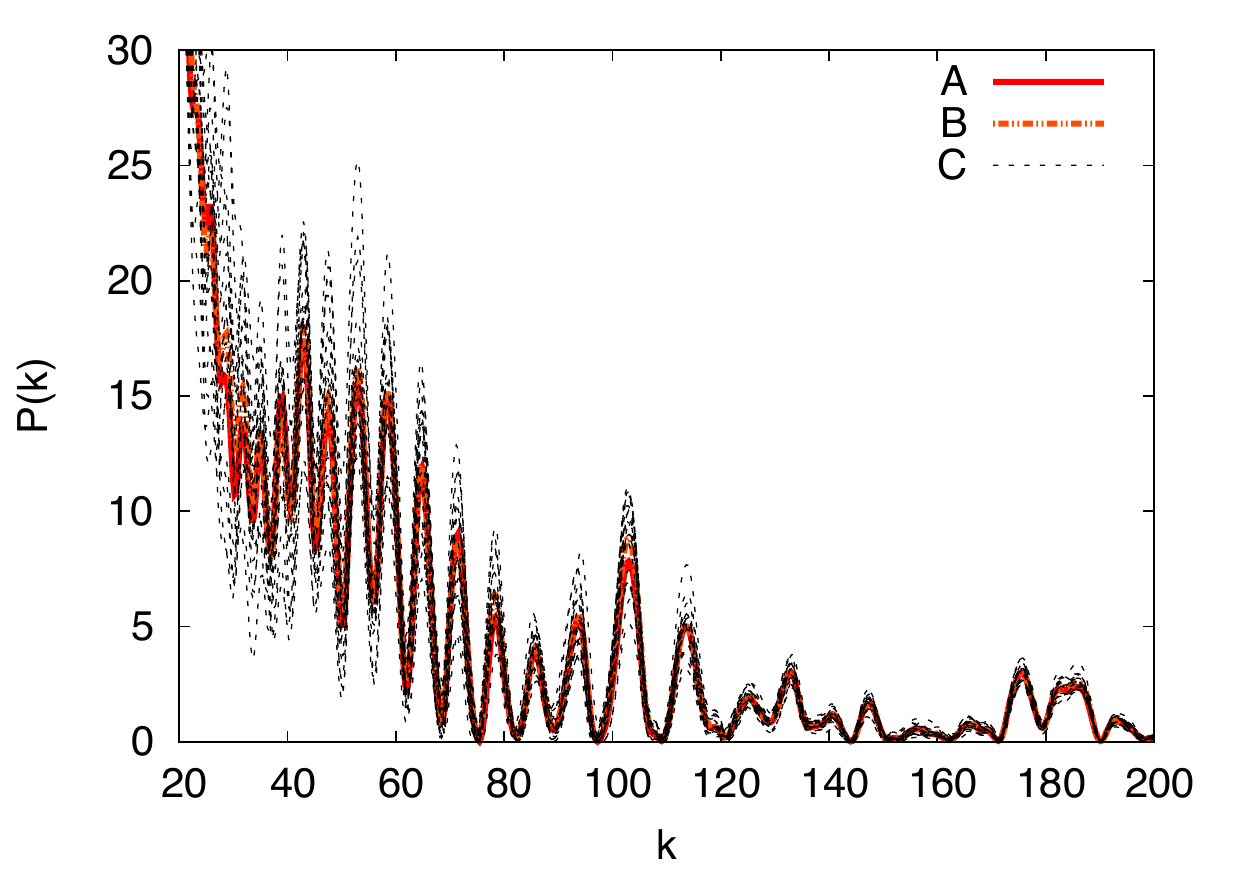}
\end{tabular}
\end{center}
\caption{ Power spectrum versus k for $L=12000$ km and  $N=8.\times 10^4$.
Upper-Left: $\Delta_\nu/E_\nu=0.20/\sqrt{E_\nu/\textrm{MeV}}+0.2 \times E_\nu/m_p$;
Upper-Right: $\Delta_\nu/E_\nu=0.20/\sqrt{E_\nu/\textrm{MeV}}+0.5 \times E_\nu/m_p$;
Lower-Left: $\Delta_\nu/E_\nu=0.20/\sqrt{E_\nu/\textrm{MeV}}+1.0 \times E_\nu/m_p$;
Lower-Right: $\Delta_\nu/E_\nu=0.20/\sqrt{E_\nu/\textrm{MeV}}+1.5 \times E_\nu/m_p$.
Line A:  theoretical expectation;
Line B:  average over 20 samples;
Line C:  20 samples with $1\sigma$ fluctuation.
Earth matter effect calculated numerically using PREM density profile with 5 layers. }
\label{figure7}
\end{figure}

In Fig. \ref{figure5} we present plots which include
1$\sigma$  fluctuation of events in each energy bin: $N_i \to N_i + \sigma_i \sqrt{N_i}$ where
$\sigma_i =0,\pm1$ which are randomly chosen.  In these plots, we present 20 samples with 20 sets of numbers $\sigma_i$.
In these plots we also show the theoretical expectation of  signal strength with no fluctuation
and the average of 20 samples. We can see that the theoretical lines agree quite well with the averages
over 20 random samples. The larger the number of events, the better these two lines agree. For $N=1.\times 10^4$,
the fluctuation is quite large as can be seen in Fig. \ref{figure5} so that the signal is basically consistent with
fluctuation. For $N=4.\times 10^4$ or $8.\times 10^8$, the fluctuation of the signal strength
is no longer that big and three peaks can be read out in the plots.  This is expected since increasing the total number 
of event increases the total statistics and hence reduces the relative fluctuation.

In Fig. \ref{figure6} we present plots with 1$\sigma$  fluctuation for a fixed number of events
but with different assumptions of energy bins. We compare cases with different values of $r_e$. One can see that
fluctuation for the case with $r_e=0.05$ is not  small. On the other hand, increasing $r_e$ to a value around $0.1\sim 0.2$ 
can significantly reduces the statistical fluctuation and increases the signal strength relative to statistical fluctuation,
although the heights of the theoretical curves and the signal strengths are not increased.
For $r_e=0.1$ or $0.2$, two peaks in these plots can be read out. 
This phenomenon is not difficult to understand. Increasing the width of energy bin increases the number of events in
each energy bin, hence reduces the statistical fluctuation in each energy bin. As long as the number of energy bins is not too small
so that the oscillation pattern can be well produced with a discrete set of energy bins, a larger value of $r_e$ 
can help to enhance the signal strength relative to statistical fluctuation. For $r_e=0.5$,  although all fluctuations tend to converge, 
the number of energy bin becomes too small so that
the oscillation pattern can not be well reproduced and there are no visible peaks above background fluctuation in this case.

In Fig. \ref{figure7} we presents plots with 1$\sigma$ fluctuation for
different assumptions of energy bins and for a fixed number of events. In this plot we compare cases with
different values of $r_a$. One can see that changing value of $r_a$ quickly changes the signal strength.
For $r_a=0.2$ or $0.5$, there are clearly two peaks with $k\approx 130$ or $180$ in the plots. 
For $r_a=1.0$, there is a peak visible at $k\approx 130$ but the peak expected at $k\approx 180$ is no longer
visible. For $r_a=1.5$, no peak is visible and everything is consistent with background fluctuation
even if the number of events is much larger, e.g. ten times larger than that in Fig. \ref{figure7}.
From these analysis, we can see that a good angular resolution with $r_a \lsim 0.5$ is very helpful
in identifying the signal of Earth matter effect in oscillation of SNe neutrinos. 

\begin{figure}[tb]
\begin{center}
\begin{tabular}{cc}
\includegraphics[scale=1,width=8cm]{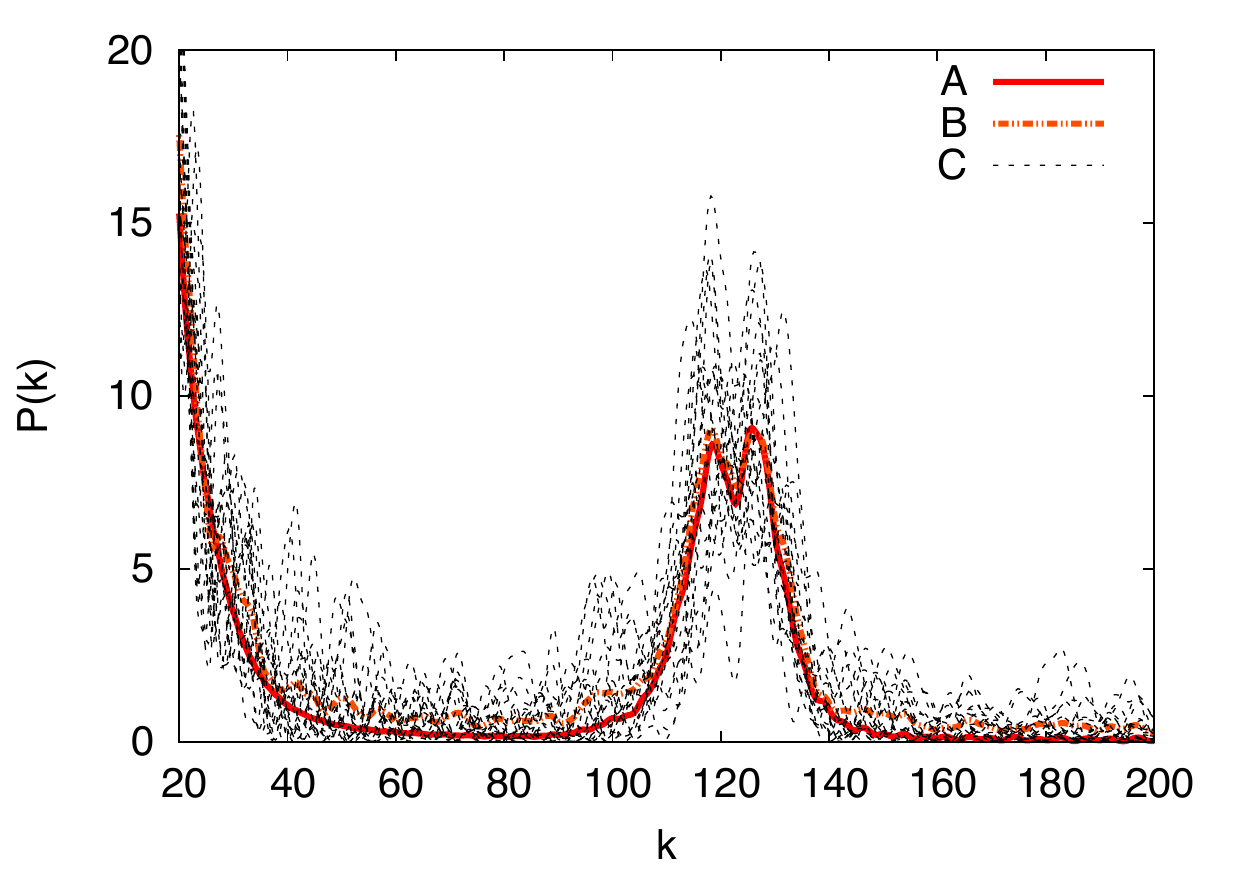}
\includegraphics[scale=1,width=8cm]{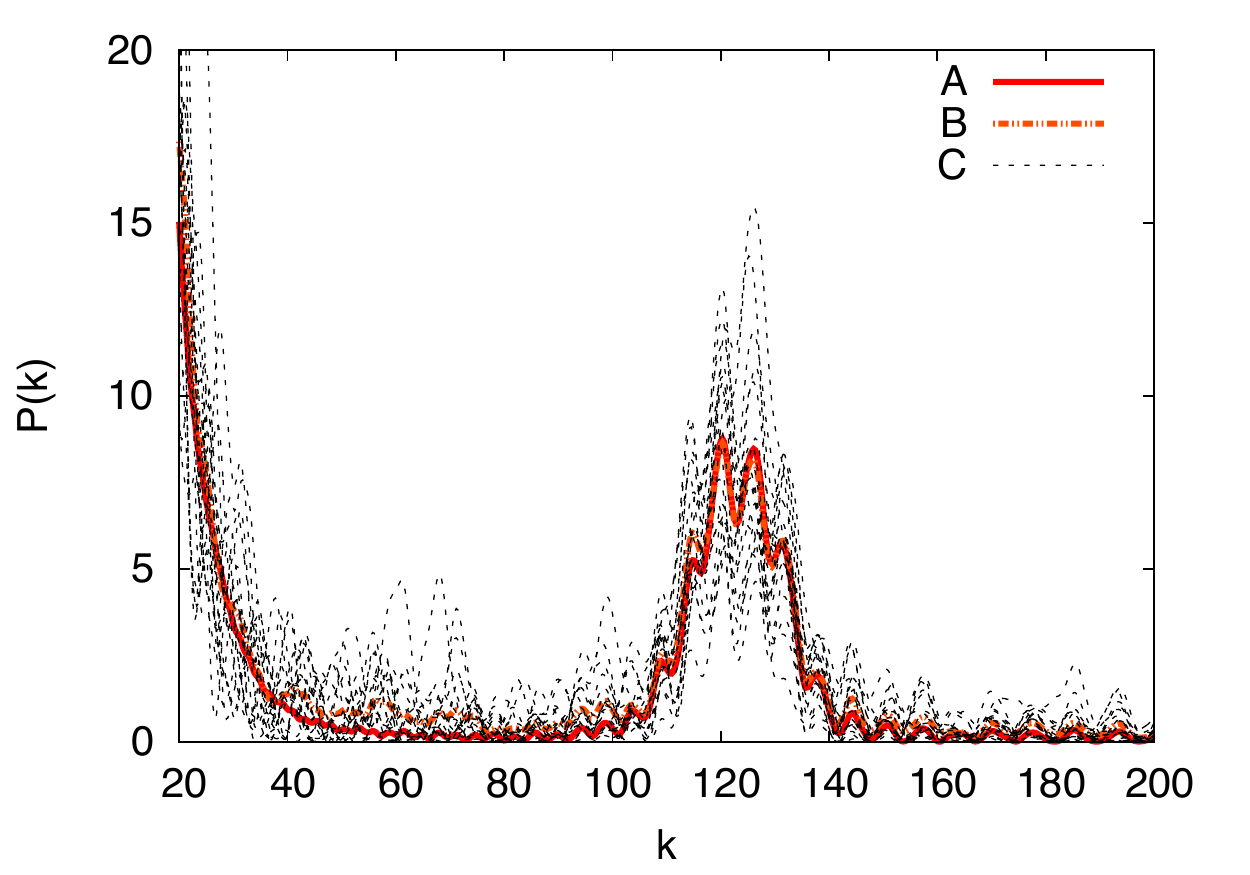}
\\
\includegraphics[scale=1,width=8cm]{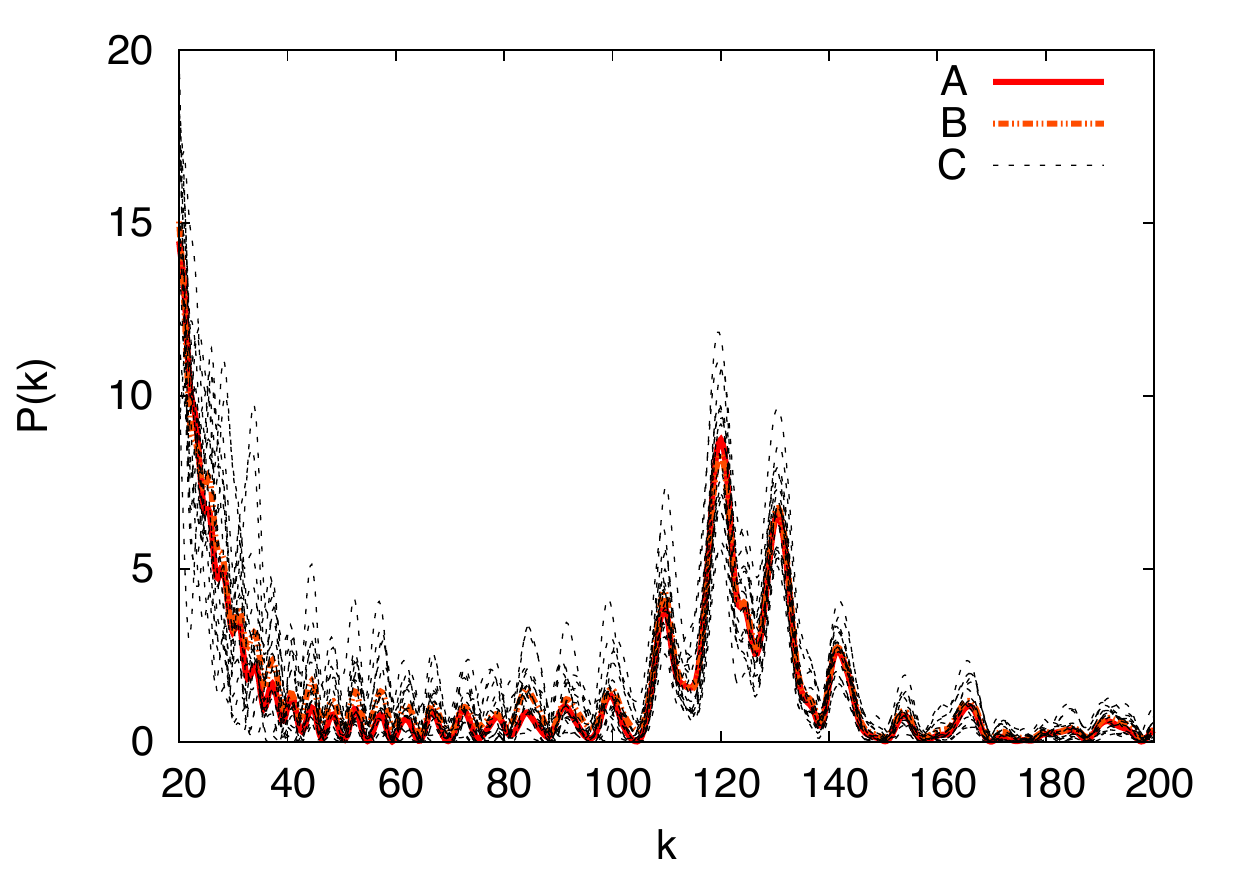}
\includegraphics[scale=1,width=8cm]{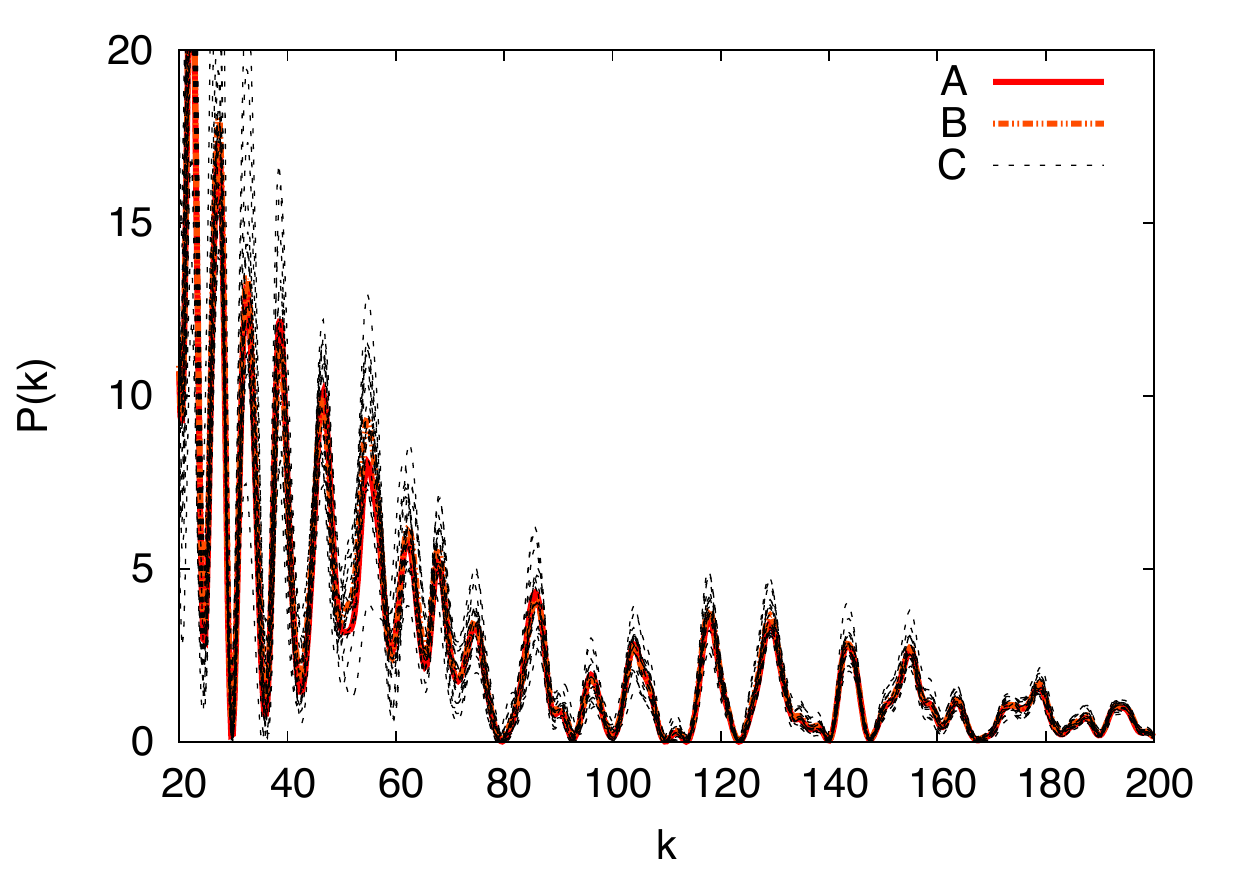}
\end{tabular}
\end{center}
\caption{ Power spectrum versus k for $L=8000$ km and  $N=4.\times 10^4$.
Upper-Left: $\Delta_\nu/E_\nu=0.05/\sqrt{E_\nu/\textrm{MeV}}+0.5 \times E_\nu/m_p$;
Upper-Right: $\Delta_\nu/E_\nu=0.10/\sqrt{E_\nu/\textrm{MeV}}+0.5 \times E_\nu/m_p$;
Lower-Left: $\Delta_\nu/E_\nu=0.20/\sqrt{E_\nu/\textrm{MeV}}+0.5 \times E_\nu/m_p$;
Lower-Right: $\Delta_\nu/E_\nu=0.50/\sqrt{E_\nu/\textrm{MeV}}+0.5 \times E_\nu/m_p$.
Line A:  theoretical expectation;
Line B:  average over 20 samples;
Line C:  20 samples with $1\sigma$ fluctuation.
Earth matter effect calculated numerically using PREM density profile with 5 layers.
}
\label{figure8}
\end{figure}

We  note that varying the function of energy bin in data analysis can provide more details of 
the spectrum of SNe neutrinos. As a comparison, in Fig. \ref{figure8}  we present plots with 
1$\sigma$  fluctuation for a fixed number of events but with different assumptions of energy bins,
similar to those in Fig. \ref{figure6} but with baseline of neutrino changed to 8000 km.
One can see that similar to plots in Fig. \ref{figure6}, increasing the widths of energy bins
reduces the statistical fluctuation and increases the signal strength relative to statistical fluctuation, as long as $r_e$ is not increased
to around $0.5$. A major difference of Fig. \ref{figure6} and Fig. \ref{figure8} is that there are two
visible peaks in Fig. \ref{figure6} and only one visible peak in Fig. \ref{figure8}. Most importantly, 
the peak in Fig. \ref{figure8} changes slowly with $r_e$ for $r_e \lsim 0.2$, similar to the peak with smaller k value in Fig. \ref{figure6}.
But the peak with larger k value in Fig. \ref{figure6} changes faster than the peak with smaller k value.
Apparently,  Fig. \ref{figure6} says that it encodes more information of the spectrum of SNe neutrinos than Fig. \ref{figure8}
and varying the width of energy bins in data analysis can reveal these informations of SNe neutrinos.
This interesting case happens for longer baseline with SNe neutrinos crossing the core of the Earth
which has three frequencies in the modulation of the spectrum of  SNe neutrinos caused by the Earth matter effect.
Apparently, to get more detailed information of the spectrum of SNe neutrinos using this method
we need a good energy resolution so that we can have more freedom to change the function of energy bins
in data analysis.

\section{Summary}
In summary, we have considered the detection of SNe neutrinos with Earth matter effect in neutrino oscillation.
As an example, we take IBD events of SNe neutrinos in our analysis.  We first show using numerical calculation that Earth matter effect
in oscillation of SNe neutrinos can be well described by an adiabatic perturbation~\cite{Liao0} which is
originally developed for describing oscillation of solar neutrinos. Then we study
detection of  SNe neutrino using Earth matter effect. Comparing with
previous works on Earth matter effect in oscillation of SNe neutrinos~\cite{SNeNeu1,SNeNeu2,LaTPC},
we analyze, in particular,
the effect of resolution of neutrino energy on this subject.  Since the reconstruction of neutrino energy
involves two aspects, i.e. the energy resolution and angular resolution of electron(positron) event in detector, 
we discuss these two aspects by approximating their effects using
two terms in (\ref{ERES1}), the function of energy bins  of neutrinos. We vary the function of energy bins and 
study the effect of energy resolution and angular resolution on detecting the Earth matter effect in oscillation
of SNe neutrinos.
 
We take into account statistical fluctuation of events in energy bins in our analysis and
study the signal strength of the Earth matter effect relative to statistical fluctuation.
 For some parameters possible to realize in the accretion phase of core-collapse SN, we show that  
an energy resolution of  positron around $(0.1\sim 0.2)/\sqrt{\textrm{$E_e/$MeV}}$ and an angular
resolution $\delta(\cos\theta) \lsim 0.5 $ can help both to suppress the statistical fluctuation and to have a reasonably
large signal strength if number of events is around a few of  ten thousands.  Such a large number of SN events can be realized for
JUNO detector with a SN around a few kpc away from the Earth~\cite{JUNO}. 

We note that analysis with other assumptions of temperatures, luminosities and initial spectra etc., can be done similarly.
Certainly, for different set of parameters, optimal choice of energy resolution and angular resolution can be different.
For example, if SN is closer and the total number of events is even larger, the width of energy bin can be smaller and meanwhile
it's still possible to suppress statistical fluctuation and has a large enough signal strength relative to statistical fluctuation.
On the other hand, if the flavor-difference of temperature is larger, the signal of Earth matter effect can show up for
smaller number of events. In this case with smaller number of events, the width of energy bin chosen in analysis can not be too small
since it's not good for suppressing the statistical fluctuation.

Since we do not really know the initial spectrum of SNe neutrinos and the relevant parameters, it's of great
virtue if the detector has a very good energy resolution. In this case, one can vary the widths of energy bins
and study the dependence of signal strength on the function of energy bin, e.g. (\ref{ERES1}). Apparently, this
provides information about the details of flavor-difference of spectra of SNe neutrinos.  We have shown that this
is in particular interesting for SNe neutrinos crossing the core of the Earth.
This is an interesting topic worth studying carefully in the future.

We emphasize that detecting the Earth matter effect in oscillation of SNe neutrinos offers an independent way 
to measure the spectrum of SNe neutrinos. For example, for the parameters used in this articles, tens of thousands
events are good enough to detect the signal of Earth matter effect which shows up the difference of $F^0_{{\bar \nu}_e}$
and $F^0_{{\bar \nu}_X}$.   We show that this requires an angular resolution of detector, at least not bad, if it's not very good. 
For LS detector, precise angular information of electron or positron is not possible to reconstruct for events with energy around tens of MeV.
However, as shown in this article, an angular resolution
at around $\delta (\cos\theta)\sim 0.5$ is already good for working out the signal of Earth matter effect in
oscillation of SNe neutrinos.  Such an angular resolution just means a bit better than identifying the 
backward or forward directions of positrons in IBD processes.  This requirement on angular resolution may not be hard to achieve. We conclude
that it might be optimistic to detect signal of Earth matter effect in oscillation of SNe neutrinos, if they are coming
from a core-collapse SN at less than around 10 kpc away from the Earth, and to detect flavor-difference
of the spectra of SNe neutrinos through detecting the Earth matter effect in oscillation of SNe neutrinos.

\acknowledgments
This work is supported by National Science Foundation of
 China(NSFC), grant No.11135009, No. 11375065 and Shanghai Key Laboratory
 of Particle Physics and Cosmology, grant No. 15DZ2272100.

\end{document}